\documentclass[12pt,titlepage,letterpaper]{utarticle}

\usepackage{amsmath}
\usepackage{mathrsfs}
\usepackage{amsfonts}
\usepackage{amssymb}
\usepackage{amsthm}
\usepackage{mathtools}
\usepackage{mathbbol}
\usepackage{graphicx}
\usepackage{color}
\usepackage{xparse}
\usepackage{tikz}
\usepackage{tocloft}
\usepackage[titletoc,title]{appendix}
\usepackage[nosort]{cite}
\usepackage{hyperref}
\usepackage{longtable}
\usepackage{comment}

\usepackage{caption}

\definecolor{lightmauve}{RGB}{255,187,255}
\definecolor{lightblue}{RGB}{238,238,255}
\definecolor{lightred}{RGB}{255,238,238}
\definecolor{midgreen}{RGB}{15,170,15}

\definecolor{aqua}{rgb}{0, 1.0, 1.0}
\definecolor{fuschia}{rgb}{1.0, 0, 1.0}
\definecolor{gray}{rgb}{0.502, 0.502, 0.502}
\definecolor{lime}{rgb}{0, 1.0, 0}
\definecolor{maroon}{rgb}{0.502, 0, 0}
\definecolor{navy}{rgb}{0, 0, 0.502}
\definecolor{olive}{rgb}{0.502, 0.502, 0}
\definecolor{purple}{rgb}{0.502, 0, 0.502}
\definecolor{silver}{rgb}{0.753, 0.753, 0.753}
\definecolor{teal}{rgb}{0, 0.502, 0.502}

\makeatletter
\newdimen\itex@wd%
\newdimen\itex@dp%
\newdimen\itex@thd%
\def\itexspace#1#2#3{\itex@wd=#3em%
\itex@wd=0.1\itex@wd%
\itex@dp=#2ex%
\itex@dp=0.1\itex@dp%
\itex@thd=#1ex%
\itex@thd=0.1\itex@thd%
\advance\itex@thd\the\itex@dp%
\makebox[\the\itex@wd]{\rule[-\the\itex@dp]{0cm}{\the\itex@thd}}}
\makeatother

\makeatletter
\newif\if@sup
\newtoks\@sups
\def\append@sup#1{\edef\act{\noexpand\@sups={\the\@sups #1}}\act}%
\def\reset@sup{\@supfalse\@sups={}}%
\def\mk@scripts#1#2{\if #2/ \if@sup ^{\the\@sups}\fi \else%
  \ifx #1_ \if@sup ^{\the\@sups}\reset@sup \fi {}_{#2}%
  \else \append@sup#2 \@suptrue \fi%
  \expandafter\mk@scripts\fi}
\def\tensor#1#2{\reset@sup#1\mk@scripts#2_/}
\def\multiscripts#1#2#3{\reset@sup{}\mk@scripts#1_/#2%
  \reset@sup\mk@scripts#3_/}
\makeatother

\makeatletter
\newbox\slashbox \setbox\slashbox=\hbox{$/$}
\def\itex@pslash#1{\setbox\@tempboxa=\hbox{$#1$}
  \@tempdima=0.5\wd\slashbox \advance\@tempdima 0.5\wd\@tempboxa
  \copy\slashbox \kern-\@tempdima \box\@tempboxa}
\def\slash{\protect\itex@pslash}
\makeatother

\def\clap#1{\hbox to 0pt{\hss#1\hss}}

\let\oldroot\root
\def\root#1#2{\oldroot #1 \of{#2}}
\renewcommand{\sqrt}[2][]{\oldroot #1 \of{#2}}

\DeclareSymbolFont{symbolsC}{U}{txsyc}{m}{n}
\SetSymbolFont{symbolsC}{bold}{U}{txsyc}{bx}{n}
\DeclareFontSubstitution{U}{txsyc}{m}{n}

\DeclareSymbolFont{stmry}{U}{stmry}{m}{n}
\SetSymbolFont{stmry}{bold}{U}{stmry}{b}{n}

\DeclareFontFamily{OMX}{MnSymbolE}{}
\DeclareSymbolFont{mnomx}{OMX}{MnSymbolE}{m}{n}
\SetSymbolFont{mnomx}{bold}{OMX}{MnSymbolE}{b}{n}
\DeclareFontShape{OMX}{MnSymbolE}{m}{n}{
    <-6>  MnSymbolE5
   <6-7>  MnSymbolE6
   <7-8>  MnSymbolE7
   <8-9>  MnSymbolE8
   <9-10> MnSymbolE9
  <10-12> MnSymbolE10
  <12->   MnSymbolE12}{}

\makeatletter
\def\re@DeclareMathSymbol#1#2#3#4{%
    \let#1=\undefined
    \DeclareMathSymbol{#1}{#2}{#3}{#4}}
\re@DeclareMathSymbol{\neArrow}{\mathrel}{symbolsC}{116}
\re@DeclareMathSymbol{\neArr}{\mathrel}{symbolsC}{116}
\re@DeclareMathSymbol{\seArrow}{\mathrel}{symbolsC}{117}
\re@DeclareMathSymbol{\seArr}{\mathrel}{symbolsC}{117}
\re@DeclareMathSymbol{\nwArrow}{\mathrel}{symbolsC}{118}
\re@DeclareMathSymbol{\nwArr}{\mathrel}{symbolsC}{118}
\re@DeclareMathSymbol{\swArrow}{\mathrel}{symbolsC}{119}
\re@DeclareMathSymbol{\swArr}{\mathrel}{symbolsC}{119}
\re@DeclareMathSymbol{\nequiv}{\mathrel}{symbolsC}{46}
\re@DeclareMathSymbol{\Perp}{\mathrel}{symbolsC}{121}
\re@DeclareMathSymbol{\Vbar}{\mathrel}{symbolsC}{121}
\re@DeclareMathSymbol{\sslash}{\mathrel}{stmry}{12}
\re@DeclareMathSymbol{\boxslash}{\mathrel}{stmry}{27}
\re@DeclareMathSymbol{\boxbslash}{\mathrel}{stmry}{28}
\re@DeclareMathSymbol{\boxast}{\mathrel}{stmry}{24}
\re@DeclareMathSymbol{\boxcircle}{\mathrel}{stmry}{29}
\re@DeclareMathSymbol{\boxbox}{\mathrel}{stmry}{30}
\re@DeclareMathSymbol{\obslash}{\mathrel}{stmry}{20}
\re@DeclareMathSymbol{\obar}{\mathrel}{stmry}{58}
\re@DeclareMathSymbol{\olessthan}{\mathrel}{stmry}{60}
\re@DeclareMathSymbol{\ogreaterthan}{\mathrel}{stmry}{61}
\re@DeclareMathSymbol{\bigsqcap}{\mathop}{stmry}{"64}
\re@DeclareMathSymbol{\biginterleave}{\mathop}{stmry}{"6}
\re@DeclareMathSymbol{\invamp}{\mathrel}{symbolsC}{77}
\re@DeclareMathSymbol{\parr}{\mathrel}{symbolsC}{77}
\makeatother

\makeatletter
\def\Decl@Mn@Delim#1#2#3#4{%
  \if\relax\noexpand#1%
    \let#1\undefined
  \fi
  \DeclareMathDelimiter{#1}{#2}{#3}{#4}{#3}{#4}}
\def\Decl@Mn@Open#1#2#3{\Decl@Mn@Delim{#1}{\mathopen}{#2}{#3}}
\def\Decl@Mn@Close#1#2#3{\Decl@Mn@Delim{#1}{\mathclose}{#2}{#3}}
\Decl@Mn@Open{\llangle}{mnomx}{'164}
\Decl@Mn@Close{\rrangle}{mnomx}{'171}
\Decl@Mn@Open{\lmoustache}{mnomx}{'245}
\Decl@Mn@Close{\rmoustache}{mnomx}{'244}
\Decl@Mn@Open{\llbracket}{stmry}{'112}
\Decl@Mn@Close{\rrbracket}{stmry}{'113}
\makeatother

\makeatletter
\DeclareRobustCommand\widecheck[1]{{\mathpalette\@widecheck{#1}}}
\def\@widecheck#1#2{%
    \setbox\z@\hbox{\m@th$#1#2$}%
    \setbox\tw@\hbox{\m@th$#1%
       \widehat{%
          \vrule\@width\z@\@height\ht\z@
          \vrule\@height\z@\@width\wd\z@}$}%
    \dp\tw@-\ht\z@
    \@tempdima\ht\z@ \advance\@tempdima2\ht\tw@ \divide\@tempdima\thr@@
    \setbox\tw@\hbox{%
       \raise\@tempdima\hbox{\scalebox{1}[-1]{\lower\@tempdima\box
\tw@}}}%
    {\ooalign{\box\tw@ \cr \box\z@}}}
\makeatother

\makeatletter
\NewDocumentCommand\mathraisebox{moom}{%
\IfNoValueTF{#2}{\def\@temp##1##2{\raisebox{#1}{$\m@th##1##2$}}}{%
\IfNoValueTF{#3}{\def\@temp##1##2{\raisebox{#1}[#2]{$\m@th##1##2$}}%
}{\def\@temp##1##2{\raisebox{#1}[#2][#3]{$\m@th##1##2$}}}}%
\mathpalette\@temp{#4}}
\makeatletter

\makeatletter
\def\udots{\mathinner{\mkern2mu\raise\p@\hbox{.}
\mkern2mu\raise4\p@\hbox{.}\mkern1mu
\raise7\p@\vbox{\kern7\p@\hbox{.}}\mkern1mu}}
\makeatother

\newcommand{\product}{\prod}

\newcommand{\widebar}{\overline}

\theoremstyle{plain}

\theoremstyle{definition}

\theoremstyle{remark}

\usetikzlibrary{positioning,positioning,arrows,decorations.markings}

\begin{document}
\renewcommand{\arraystretch}{1.5}

\preprint{
UTWI-29-2024\\
}

\title{Non-Simply Laced Class-S Vertex Operator Algebras}

\author{Grant Elliot}
     \oneaddress{
      Weinberg Institute for Theoretical Physics\\
      Department of Physics,\\
      University of Texas at Austin,\\
      Austin, TX 78712, USA \\
      {~}\\
      \email{gelliot123@utexas.edu}
      }

\date{September 30, 2024}

\Abstract{In \cite{Arakawa:2018egx} the VOAs associated to 4d $\mathcal{N}=2$ class-S theories were constructed in addition to a generalization for non-simply laced Lie algebras. However, 6d (2,0) theories have an ADE classification, and therefore class-S theories which are engineered by them come in ADE types. Thus, these non-simply laced VOAs are not thought to correspond to 4d physical theories. Regardless, we analyze these VOAs and their Drinfeld-Sokolov reductions in an effort to determine their properties. We find that many of these reduced VOAs appear to correspond to actual 4d $\mathcal{N}=2$ SCFTs. Additionally, we find what appear to be $F_4$ instanton VOAs, and hence from the proposal of \cite{Beem:2022mde} these correspond to 3d $\mathcal{N}=4$ quiver gauge theories whose Coulomb branches are $F_4$ instanton moduli spaces. Using a construction of twisted class-S VOAs from non-simply laced ones, we find additional evidence that the $F_4$ instanton VOAs do not correspond to four-dimensional field theories and outline an analogous argument for the $G_2$ instanton case. Our method appears to be quite general and may be a promising technique for ruling out 4d origins of VOAs that are similar to those of 4d $\mathcal{N}=2$ SCFTs, in addition to constructing VOAs of actual 4d theories.
}

\maketitle

\tocloftpagestyle{empty}
\tableofcontents
\vfill
\newpage
\setcounter{page}{1}

\section{Introduction}\label{introduction}
Class-S theories \cite{Gaiotto:2009we,Gaiotto:2009hg} arise from the compactification of a six-dimensional $\mathcal{N} =(2,0)$ theory on a Riemann surface with a partial topological twist preserving supersymmetry. The 6d $\mathcal{N}=(2,0)$ SCFTs originate from compactifying type IIB string theory on an ALE space and thus have an ADE classification, as can also be shown from anomaly cancellation \cite{Henningson:2004dh}. Hence, class-S theories come in ADE types, which can additionally be supplemented by adding twist lines going around cycles on the Riemann surface. In the IIB picture this corresponds to a monodromy of the ALE space.

What makes class-S theories so appealing and first demonstrated in \cite{Alday:2009aq} is that many of their features can be understood from a theory on the Riemann surface. Since the $(2,0) $ theory is conformal, compactifying on either the Riemann surface or four-manifold first should yield the same result. Additionally, if there is some natural object assigned to a class-S theory which does not depend on the conformal manifold, there should exist a topological field theory valued in said object, which occurs for instance in the case of the superconformal index\cite{Gadde:2011ik,Mekareeya:2012tn} and Higgs branch\cite{Moore:2011ee}.

One of the most refined invariants of a 4d $\mathcal{N}=2$ SCFT is the associated VOA of \cite{Beem:2013sza}, which for class-S were discussed in the physics literature in \cite{Beem:2014rza} and constructed rigorously in \cite{Arakawa:2018egx}. Arakawa's construction is equally valid for non-simply laced groups in alignment with previous work on their Higgs branches \cite{Ginzberg,Braverman:2017ofm}. 

The VOAs constructed by Arakawa are specified by a simple Lie algebra $\mathfrak{j}$ and a Riemann surface with $n$ marked points, and will contain $n$ copies of the affine $\mathfrak{j}$ current algebra at the critical level. The case of a sphere with three marked points is referred to as a trinion. None of the non-simply laced trinion VOAs correspond to known 4d $\mathcal{N}=2$ SCFTs. However, we can perform Drinfeld-Sokolov reduction\cite{MR2013802} on these VOAs to obtain new VOAs, which are the main subject of study of this paper. Thus, the most general (untwisted) class-S VOA is then determined by the data of a simple Lie algebra $\mathfrak{j}$ and a Riemann surface with $n$ marked points labeled by nilpotent orbits of $\mathfrak{j}$. The simply laced case coincides with the VOA of the 4d theory. Surprisingly, in the non-simply laced case we will find many of the reduced VOAs correspond to known 4d SCFTs, and the most notable VOAs not corresponding to known 4d theories we find are $F_4$ instanton VOAs. While we do not compare these directly via OPEs, in the rank one and two cases we find matching current algebra levels, central charges and characters to high orders in $\tau$. Furthermore, we check that the VOAs behave as expected under Drinfeld-Sokolov reduction.  

Additionally, we can formally carry over the Hall-Littlewood index and attempt to calculate the Hilbert series of the associated variety \cite{arakawa2012remark} of these VOAs. We expect this expression to come from the 3d perspective, which we detail later. We find perfect agreement in all examples we test, such as the rank-one $F_4$ instanton theory. 

Amongst the VOAs that we find that seem to correspond to 4d $\mathcal{N}=2$ SCFTs, there is a clear pattern of equivalence for certain VOAs. Using the notation from earlier, the twisted theories of type $\mathfrak{j}\neq A_{2n}$ with two twisted full punctures and $n$ untwisted simple punctures is isomorphic to the type $\mathfrak{g}$ class-S VOA with two full punctures and $n$ subregular punctures. There is a slightly stronger statement in the twisted $D_n$ case which we detail later.

These equivalences raise all the more questions on why many of these VOAs don't seem to correspond to 4d theories. In some cases there are issues with anomalies. Additionally we find various dualities with twisted class-S VOAs that disprove the existence of a corresponding 4d theory for many examples, such as the $F_4$ instanton VOAs. We find evidence for similar dualities completely outside of the class-S case, and use these to argue against the existence of $G_2$ instanton theories. These dualities involve a gauging of the current algebras of these VOAs, and result in the VOA of an actual 4d $\mathcal{N}=2$ SCFT whose conformal manifold has dimension incompatible with the would-be duality. In fact, we can often construct VOAs of isolated SCFTs via these gaugings. This phenomenon seems somewhat common, and we find that three rank-one theories have VOAs that can be constructed in this manner.

We give an overview of class-S VOAs and then propose the generic isomorphisms in Section \ref{Review}. In Section \ref{Instantons} we provide a brief review of instanton moduli spaces and VOAs. We then begin our analysis of class-S VOAs with the simplest non-simply laced Lie algebra, $C_2=B_2$ and then look at the $C_3,G_2$ and $B_3$ cases in Section \ref{Catalog}. We go on to discuss the Hilbert series of their associated varieties and aspects of the corresponding 3d quiver gauge theories in Section \ref{Quivers}. Lastly, we discuss various dualities involving these non-simply laced VOAs and twisted class-S VOAs and how this rules out many of the non-simply laced ones from having a 4d origin in Section \ref{Dualities}.

Due to our interest in 4d $\mathcal{N}=2$ SCFTs, we use the 4d levels for current algebras primarily throughout the text, which are related to 2d levels by $k_{2d}=-\frac{1}{2}k_{4d}$. This is quite convenient when we relate the non-simply laced VOAs to the VOAs of known 4d theories. We additionally use various vocabulary from the 4d literature. A fixture just means the VOA obtained by performing various Drinfeld-Sokolov reductions of the three current algebras of the Class-S VOA trinion. Hence, a fixture is specified by a triple of nilpotent orbits, and a puncture corresponds to one of the nilpotent orbits. We also note that we do not always distinguish between a nilpotent orbit and its closure, as it is typically clear which we are referring to based on the context.

\section{Class-S Review and Isomorphisms}\label{Review}
\subsection{Review}

As mentioned in the introduction, class-S theories are engineered by compactifying (2,0) theories on a Riemann surface. One can additionally place codimension-two defects of the 6D theory at points on the Riemann surface. For type $\mathfrak{j}$ ADE (2,0) theory, the defects are labeled by nilpotent orbits\footnote{The Jacobson-Morozov theorem states that these nilpotent orbits are in one to one correspondence with homomorphisms of $\mathfrak{su}(2)$ into $\mathfrak{j}$ up to conjugacy. For classical Lie algebras, these are typically specified by a partition describing how the fundamental representation decomposes under the homomorphism. See \cite{CollingwoodMcGovern} for an introduction to the subject.} of the corresponding complex Lie algebra \cite{Chacaltana:2012zy}. When we compactify with a twist line, there can be a twist going around a defect, and such defects are labeled by nilpotent orbits of the Langlands dual of the invariant subalgebra of the outer automorphism $\mathfrak{g}$. We note there exist other codimension-two defects, called wild, but we will not discuss them here.

Let us provide a quick summary of Arakawa's construction\cite{Arakawa:2018egx}, see also \cite{Beem:2022mde} for an explanation for physicists. For any simple Lie algebra $\mathfrak{j}$, the process begins with a VOA called the $\mathfrak{j}$ affine equivariant W-algebra, which we denote $\mathbf{W}_{J}$. This contains an affine current algebra of type $\mathfrak{j}$ at the critical level\footnote{The critical level is given by $k_{2d} =-\frac{1}{2}k_{4d}=-h^{\vee}$.}, which itself contains a center called the Feigin-Frenkel center. Tensoring $n$ of these $\mathbf{W}_J$ together, then identifying their Feigen-Frenkel centers via BRST cohomology, gives the VOA of a genus zero class-S VOA with $n$ full punctures. Higher genus VOAs are obtained via a BRST gauging of the diagonal current algebras associated to two full punctures. This same procedure can be used for non-simply laced Lie algebras, however there does not appear to be a corresponding 4d SCFT. There is also a mixed Feigin-Frenkel gluing \cite{Beem:2022mde}, which enables the construction of the VOAs of twisted class-S theories. We will discuss this construction in slightly more detail in Subsection \ref{TwistedTrinion} where we will make extensive use of it. 

Performing a Drinfeld-Sokolov reduction corresponding to a nilpotent orbit of $\mathfrak{j}$ or $\mathfrak{g}$ of the current algebra associated to one of the punctures gives the VOA of a class-S theory whose full puncture has been replaced by a non-full puncture. On the 4d side this corresponds to nilpotent Higgsing \cite{Beem:2014rza}. 

The classification of three punctured spheres is particularly important as all other class-S theories/VOAs can be obtained by gluing them together. For ordinary class-S theories, this has been mostly worked out in a series of papers\cite{Chacaltana:2010ks,Chacaltana:2011ze,Chacaltana:2012ch,Chacaltana:2013oka,Chacaltana:2014jba,Chacaltana:2015bna,Chacaltana:2016shw,Chacaltana:2017boe,Chacaltana:2018vhp,Distler:2021cwz}. Classifying the low rank non-simply laced VOAs corresponding to three punctured spheres is one of the main motivations of this paper. However, we restrict to VOAs which are not bad, that is the unrefined character does not diverge. 

Compactifying a class-S theory on a circle gives a 3d $\mathcal{N}=4$ SCFT. This turns out to have a 3d mirror given by a star shaped quiver \cite{Benini:2010uu}. The central node will be the $G^{\vee}$, and the quivers glued to it are determined by the punctures. The Coulomb branch of the glued quivers will be the nilpotent Slodowy slice of the corresponding orbit, while the Higgs branch will be a possibly trivial cover of the nilpotent orbit $d(\mathcal{O})$\footnote{Here $d$ denotes the Spaltenstein map\cite{BarbaschVogan,Spaltenstein}.}. When the cover is trivial, this is just the $T_{\rho}(G)$ theory discussed in \cite{Gaiotto:2008ak}. For a genus $g$ theory we additionally add $g$ adjoint hypermultiplets charged under the central node. If the theory is twisted, the presence of four or more twisted punctures adds more hypers charged under the central node. 

The Coulomb branch of this quiver is the Higgs branch of the 4d theory, and hence the associated variety of the VOA. While the non-simply laced class-S VOAs have no 4d interpretation, in \cite{Beem:2022mde} they were conjectured to be the C-twist boundary vertex operator algebras \cite{Costello:2018fnz} of the star shaped quiver that are the same as the twisted ones, but without the extra hypers coming from the twisted punctures charged under the central node. Hence, the associated varieties of the VOAs should be the Coulomb branch of the corresponding quiver. 

We will remain somewhat agnostic about this conjecture throughout the paper, though we will assume the weaker statement that the associated varieties of the VOAs are the corresponding Coulomb branches. As for the exact nature of the VOAs, it is not clear that they are the C-twist boundary VOAs of these 3d theories. There are examples of 4d theories whose 3d mirror C-twist VOAs do not agree with the VOAs of the 4d theories, such as the $(A_1,A_3)$ Argyres-Douglas theory as noted in \cite{Yoshida:2023wyt}. Another example is the $(A_1,D_4)$ AD theory, whose 4d VOA is a $\mathfrak{su}(3)$ current algebra at 2d level $-\frac{3}{2}$ which is not the H-twist 3d boundary VOA of the circle compactification \cite{Beem:2023dub}. This is notably a regular class-S theory\cite{Beem:2020pry}, albeit a twisted one, however the point stands that there is no a priori reason to think the H-twisted boundary VOA should agree with the VOA of the parent 4d theory for general class-S VOAs. Regardless, we find significant evidence that these non-simply laced class-S VOAs are some invariant of the 3d theory, which in the case of 3d SCFTs arising from circle compactifications of 4d SCFTs, coincides with the 4d VOA. In Section \ref{Discuss}, we discuss the possibility that these VOAs are obtained by applying the free field realizations of \cite{Beem:2019tfp} to the Coulomb branches of these quivers.

\subsection{Isomorphisms}
There turns out to be a clear pattern of isomorphisms of these non-simply laced class-S VOAs with certain twisted class-S VOAs. Let $\mathfrak{j} \neq A_{2n}$ be an ADE Lie algebra with $\mathbb{Z}_2$ outer-automorphism and let $\mathfrak{g}$ be the Langlands dual of the Lie algebra invariant under the automorphism. Let $\mathcal{O}$ and $\mathcal{O}'$ denote the subregular orbits of $\mathfrak{j}$ and $\mathfrak{g}$ respectively. We conjecture the following VOAs to be equivalent: the twisted class-S VOA of type $\mathfrak{j}$ with two twisted full punctures and $n$ $\mathcal{O}$ untwisted punctures; the type $\mathfrak{g}$ class-S VOA with two full punctures and $n$ $\mathcal{O}'$ punctures. This implies many other isomorphisms of VOAs obtained by gluing the full punctures and/or performing various Drinfeld-Sokolov reductions. 

We can actually make a stronger statement, that is pairs of $\mathcal{O}$ and $\mathcal{O}'$ result in the same equivalence as long as the nilpotent Slodowy slice of $\mathcal{O}$ and $\mathcal{O}'$ are the same\footnote{In the $D_n$ case, there are very even orbits which are described by the same partition of $2n$ and have the same nilpotent Slodowy slice, but in the language of \cite{Chacaltana:2013oka} are distinguished by their color. It turns out that both orbits are paired with the same $\mathcal{O}'$. This is not a problem, as the choice of very even orbit does not affect the VOA in the twisted $D_n$ case. To see this, note that in the twisted $D_n$ case there is a $\mathbb{Z}_{2}$ symmetry that changes the color of every very even orbit. Additionally, the color of a very even orbit can be changed by moving it around a twisted puncture. This process only involves moving around in the conformal manifold, and thus does not affect the VOA.}. This only affects the twisted $D_n$ case, as the nilpotent Slodowy slices corresponding to the $D_{n+1}$ partition $\mathcal{S}_{2n-k+1,k+1}$ and the $C_n$ partition $\mathcal{S}_{2n-k,k}$ are isomorphic\cite{Henderson_2014}. Additionally, the nilpotent Slodowy slice of the $G_2$ nilpotent orbit $\tilde{A_1}$ is isomorphic to the nilpotent Slodowy slice of the $C_3$ orbit $[4,1^2]$, which results in certain isomorphisms in the $S_3$ twisted $D_4$ theory.

\begin{table}
    \centering
    \begin{tabular}{|c|c|c|} \hline 
        $n$ & $D_n$ punctures  & $C_{n-1}$ punctures \\ \hline 
        4 & $[5,3]$ & $[4,2]$\\
        & $[4^2]$ & $[3^2]$\\
        \hline
        5 & $[7,3]$ & $[6,2]$\\ 
        & $[5^2]$ & $[4^2]$\\ \hline  
        6 &  $[9,3]$& $[8,2]$\\ 
        &  $[6^2]$& $[5^2]$\\
 & $[7,5]$&$[6,4]$\\ \hline
    \end{tabular}
    \caption{Some examples of pairs of punctures that appear to result in isomorphic VOAs.}
    \label{tab:my_label}
\end{table}

We perform a variety of consistency checks for these proposed isomorphisms. First, one can show that the central charges and current algebras are the same. Second, at the level of associated varieties, the effect of removing a simple untwisted $\mathfrak{j}$ puncture and a simple $\mathfrak{g}$ puncture are the same since the nilpotent Slodowy slices of the two corresponding subregular orbits are the same. Third, the global forms of the flavour symmetry are the same \cite{Distler:2022nsn}. Lastly, we compare the corresponding characters and find agreement to very high orders in $\tau$ in a large number of examples.

We also note that at the level of indices, the equivalence of characters seems to follow from\footnote{Both characters are sums over representations of $\mathfrak{g}$, with the choice of either the L.H.S. or R.H.S. of Equation \ref{Character} as a factor being the only difference in the formulas.} 
\begin{equation}
\label{Character}
    \frac{\psi_{\mathfrak{R}}^{\rho}(\mathbf{a}_i,\tau)}{\psi_{\mathfrak{R}}^{\rho_{f}}(\tau)} = \frac{\tilde{\psi}_{\mathfrak{R}}^{\rho'}(\mathbf{a}_i,\tau)}{\tilde{\psi}_{\mathfrak{R}}^{\rho'_{f}}(\tau)}.
\end{equation}
Here $\rho$ and $\rho'$ denote the corresponding homomorphisms for $\mathcal{O}$ and $\mathcal{O}'$, while $\psi_{\mathfrak{R}}$ and $\tilde\psi_{\mathfrak{R}}$ denote the wavefunctions for Lie algebras $\mathfrak{j}$ and $\mathfrak{g}$ respectively. See Section \ref{Catalog} for more details on the characters of class-S VOAs. We have checked the above equation for some very large representations in a variety of examples, but we know no proof. The Plethystic exponentials in the K-factors cancel quite nicely, but the cancellation of the characters or Hall-Littlewood polynomials is mysterious to the author.

We conjecture this equivalence of VOAs can be explained in terms of 3d $\mathcal{N}=4$ SCFTs \cite{Costello:2018fnz}, assuming the non-simply laced class-S VOAs are invariants of the 3d theories. In \cite{Beem:2022mde}, it was suggested that the non-simply laced class-S VOAs corresponded to certain star shaped quivers resembling the mirrors of class-S theories without the additional hypers coming from twisted punctures. Consider such a mirror with two $T[G]$ quivers glued to a central node as well as the quiver for the $\mathcal{O}'$ orbit. This $\mathcal{O}'$ quiver should have Coulomb branch equal to the nilpotent Slodowy slice of $\mathcal{O}'$ and Higgs branch equal to a cover of $d(\mathcal{O}')$. 

We could similarly attach the mirror quiver for the $\mathcal{O}$ orbit of $\mathfrak{j}$, which has the same Coulomb branch as above and its Higgs branch is the orbit $d(\mathcal{O})$ of $\mathfrak{j}$. We argue that this nilpotent orbit of $\mathfrak{j}$ is precisely the aforementioned cover of the nilpotent orbit of $\mathfrak{g}$, and that the quivers attached to the central node are actually the same. It often happens that a certain nilpotent orbit in a Lie algebra is a cover of another nilpotent orbit in another algebra \cite{MR1239505}. Fortunately, when $\mathcal{O}$ and $\mathcal{O'}$ have the same nilpotent Slodowy slice, the orbit $d(\mathcal{O})$ of $\mathfrak{j}$ is a cover of the orbit $d(\mathcal{O}')$ of $\mathfrak{g}^{\vee}$, see Table \ref{Shared}. 

\begin{table}[ht!]
    \centering
    \begin{tabular}{|c|c|c|c|c|c|c|} \hline
         $\mathfrak{j}$&  $\mathcal{O}$&  $d(\mathcal{O})$&  $\mathfrak{g}$& $\mathcal{O}'$& $\mathfrak{g}^{\vee}$& $d(\mathcal{O}')$ \\ \hline
         $D_n$& $[2n-r-2,r+2]$ & $[2^{2r},1^{2n-4r}]$ &  $C_{n-1}$& $[2n-3-r,r+1]$ & $B_{n-1}$& $[3,2^{2r-2},1^{2n-4r}]$\\ \hline
         $A_{2n-1}$& $[2n-1,1]$ & $[2,1^{2n-2}]$ &  $B_{n-1}$& $[2n-2,2]$ & $C_{n-1}$ &$[2^2,1^{2n-4}]$ \\ \hline
 $E_6$& $E_6(a_1)$ &$A_1$ & $F_4$& $F_4(a_1)$& $F_4$& $\tilde{A_1}$\\ \hline
    \end{tabular}
    \caption{Isomorphic nilpotent Slodowy slices and shared nilpotent orbits in the $\mathbb{Z}_2$ twisted case. Here $d(\mathcal{O})$ is a two-fold cover of $d(\mathcal{O}')$. Note that $r$ must be chosen so that the corresponding partitions are of the correct type. While $r$ is typically odd, there is always one choice where $r$ is even. In this case, the D-partition describes two very even orbits.}
    \label{Shared}
\end{table}

This suggests the quivers attached to the central node for the subregular nilpotent orbits of $\mathfrak{g}$ and $\mathfrak{j}$, are the same. Hence, we expect that $T_{\mathcal{O}}(J)$ and $T_{\mathcal{O}'}(G)$ are related by a discrete gauging. This can be seen explicitly in the twisted $A_{2n-1}$ case as shown in Figure \ref{Discrete}  while in the twisted $D_{n}$ case it is easier to see the discrete gauging relation via their mirrors \cite{Cabrera:2017ucb}.
\begin{figure}[ht!]
\centering
\begin{tikzpicture}
\node (0) at (0,0) {$SO(2)$};
\node (A1) at (2.5,0) {$[SU(2n)]$};
\node (a) at (0,-1.5) {$O(2)$};
\node (b) at (2.5,-1.5) {$[Sp(n)]$};
\path[thick,draw,color=black] 
(0) edge  (A1)
(a) edge (b)
;
\end{tikzpicture}
\caption{The top quiver is $T_{[n-1,1]}(A_{2n-1})$ while the bottom one is $T_{[2n-1,1^2]}(B_{n})$. The flavour symmetry is indicated by square brackets.  They differ by replacing a special orthogonal factor with an orthogonal one. The top quiver is glued to the central node for both subregular Nahm orbits in the twisted $A_{2n-1}$ theory.}
\label{Discrete}
\end{figure}

\subsection{Additional Consequences}

Here we detail some additional consequences of the above observations for ordinary class-S VOAs, that will turn out to be relevant for Section \ref{Dualities}. Let $\mathcal{O}$ and $\mathcal{O'}$ as before have the same nilpotent Slodowy slice. Then the twisted $\mathfrak{j}$ theory with two $\mathcal{O}$ untwisted punctures and two twisted regular punctures should be isomorphic to the theory with two twisted $\mathcal{O}'$ punctures. At the level of characters the equivalence follows from Equation \ref{Character} since it implies
\begin{equation}\label{atypical}
    \frac{(\psi_{\mathfrak{R}}^{\rho}(\mathbf{a}_i,\tau))^2(\tilde{\psi}_{\mathfrak{R}}^{\rho'_{f}}(\tau))^2}{(\psi_{\mathfrak{R}}^{\rho_{f}}(\tau))^4} = \frac{(\tilde{\psi}_{\mathfrak{R}}^{\rho'}(\mathbf{a}_i,\tau))^2}{(\psi_{\mathfrak{R}}^{\rho_{f}}(\tau))^2}.
\end{equation}

At the 4d level, we expect there to be an irregular fixture for the OPE of the twisted regular orbit puncture and $\mathcal{O}$ that results in a trivial gauging of the $\mathcal{O}'$ puncture. In fact, for precisely these pairs with the same nilpotent Slodowy slice this property was found in \cite{Chacaltana:2012ch,Chacaltana:2013oka,Chacaltana:2015bna,Chacaltana:2016shw,Distler:2021cwz}. In this case the puncture $\mathcal{O}'$ is called atypical, and note that a fixture with an atypical puncture has a non-trivial conformal manifold. We say that the puncture $\mathcal{O}'$ resolves into the punctures $\mathcal{O}$ and the twisted regular orbit puncture. We could take Equation \ref{atypical} as a given from the 4d perspective, and then derive Equation \ref{Character} by taking the square root of both sides.

Another consequence is that a theory with both a $\mathcal{O}$ and $\mathcal{O}'$ puncture should have a $\mathbb{Z}_2$ automorphism symmetry. Hence, if we perform a Higgsing that results in some puncture replacement  $\mathcal{O}\to \tilde{\mathcal{O}}$ we could alternatively do the same Higgsing for $\mathcal{O}' \to \tilde{\mathcal{O}}'$ and we should get the same theory. From the 4d perspective we expect the OPE of $\mathcal{O}$ with $\tilde{\mathcal{O}}'$ should be equal to the OPE of $\mathcal{O}'$ with $\tilde{\mathcal{O}}$. This can also be confirmed in a variety of the twisted tinkertoy papers. As an example, the irregular fixture obtained from colliding the $[4^2]$ and $[4,2]$ punctures and the one obtained by $[5,3]$ and $[3^2]$ punctures are the same in the twisted $D_4$ theory.

\section{Instanton Moduli Spaces and VOAs}\label{Instantons}

We give a brief review of instanton moduli spaces and the corresponding VOAs, see \cite{Beem:2019snk} for a more substantial discussion. The ADHM \cite{Atiyah:1978ri} construction allows one to understand the instanton moduli space for classical Lie algebras as a hyperkh{\"a}ler quotient. Physically one can find a Lagrangian field theory whose Higgs branch is the corresponding instanton moduli space. The exceptional case is not so simple, but one can still engineer field theories whose Higgs branches are the moduli space of E-type instantons via D3 branes probing a certain F-theory singularity. The E-type cases have realizations in class-S, which allows one to obtain the Hilbert series for their Higgs branches \cite{Gaiotto:2012uq}.

Another tool one can use is 3d mirror symmetry to arrive at a Lagrangian theory whose Coulomb branch is the Higgs branch of a non-Lagrangian theory, which in our case we will take to be the moduli space of instantons. One can generalize the notion of a quiver to non-simply laced quivers, though the physical interpretation of such quivers is less clear. From these non-simply laced quivers one can obtain the moduli space of $F_4$ and $G_2$ instantons \cite{Cremonesi:2014xha}.

Since we are mainly concerned with 4d $\mathcal{N}=2$ SCFTs, we will draw attention to the fact that one can engineer a field theory whose Higgs branch is the $n$ centered moduli space of either $A_0,A_1,A_2,D_4,E_6,E_7,E_8$ instantons via $n$ D3 branes probing the corresponding F-theory singularity\cite{Sen:1996vd,Banks:1996nj,Dasgupta:1996ij,Minahan:1996cj,Minahan:1996fg}. In the case of a single $D_3$ brane these are referred to as the rank-one instanton SCFTs. What is notable about such theories is they simultaneously saturate three 4d unitarity bounds\cite{Beem:2013sza,Lemos:2015orc,Beem:2017ooy}. Excluding $A_0$, the VOAs associated to the above theories are the corresponding current algebras at 2d level $\frac{-h^{\vee}-6}{6}$. Somewhat suggestively, the Deligne-Cvitanovi\'c (DC) exceptional series\cite{Deligne,Cvitanovic:2008zz} contains the above mentioned Lie algebras as well as two others, $G_2$ and $F_4$:

\begin{equation*}
A_0,A_1,A_2,G_2,D_4,F_4,E_6,E_7,E_8.
\end{equation*}

Curiously, if these $F_4$ or $G_2$ instanton theories existed they would also saturate the known unitarity bounds. However, the rank-one theories would naively have a single CB parameter with scaling dimension $\frac{5}{2}$ and $\frac{10}{3}$ respectively, and notably such scaling dimensions don't occur in the allowed values for rank-one SCFTs. 

There are additional issues with the $F_4$ case, as if such a theory were to exist, then anomaly matching with the IR theory on its Higgs branch reveals that the $Sp(3)$ subgroup of $F_4$ would need to carry Witten's global $\mathbb{Z}_2$ anomaly\cite{Witten:1982fp}. However, this should not be possible since it is embedded in $F_4$ as argued in \cite{Shimizu:2017kzs}.

Nevertheless it was found that due to the special representation theory of the DC series one can uniformly construct the rank-two instanton VOAs even in the $F_4$ and $G_2$ case \cite{Beem:2019snk}. Quite satisfyingly we will find  both the rank-two and rank-one $F_4$ instanton VOAs as non-simply laced class-S VOAs. From the proposal of \cite{Beem:2022mde} this suggests that certain star shaped quivers have the corresponding $F_4$ instanton moduli spaces as their Coulomb branches. However, we will find various properties of the $F_4$ instanton VOAs made manifest by their class-S realizations show they cannot be the VOAs of four-dimensional $\mathcal{N}=2$ superconformal field theories. That is not to say these VOAs have nothing to do with four-dimensional physics, as we find various VOAs of known 4d theories arise as BRST reductions of $F_4$ instanton VOAs, and evidence for similar constructions involving $G_2$ instanton VOAs.

\section{Catalog}\label{Catalog}

We now begin our analysis of the VOAs. Rather than compute the OPEs explicitly, we will primarily rely on other objects in order to understand them. We can compute their central charges, and using the characters, we can compute their flavour symmetry/current algebra. The character in \cite{Arakawa:2018egx} for a three punctured sphere with three full punctures was shown in \cite{Beem:2022mde} to be equivalent to
\begin{equation}
    \mathcal{I}_{\text{Schur}} = \sum_{\mathfrak{R}}\frac{\product_{i=1}^3\psi^{0}_{\mathfrak{R}}(\mathbf{a}_i,\tau)}{\psi^{\rho}_{\mathfrak{R}}(\tau)}
\end{equation}
where the sum is over irreducible representations of the corresponding Lie algebra. Additionally, we have
\begin{equation}
 \psi^{0}_{\mathfrak{R}}(\mathbf{a}_i,\tau) = \text{P.E.}\Bigg[\frac{\chi_{\text{adj}}(\mathbf{a}_i)\tau^2}{1-\tau^2}\Bigg]\chi_{\mathfrak{R}}(\mathbf{a}_i) 
\end{equation}
where $\chi_{\mathfrak{R}}(\mathbf{a}_i)$ denotes the character of the corresponding representation, P.E. is the Plethystic exponential and
\begin{equation}
 \psi^{\rho}_{\mathfrak{R}}(\mathbf{a}_i,\tau) = \text{P.E.}\Bigg[\frac{\sum_{d}\tau^{2d}}{1-\tau^2}\Bigg]\chi_{\mathfrak{R}}^{\rho}(\tau) .
\end{equation}
The sum is over the degrees of the corresponding Lie algebra and $\chi_{\mathfrak{R}}^{\rho}(\tau)$ denotes the character of the corresponding representation of $\mathfrak{su}(2)$ obtained by decomposing the representation of $\mathfrak{j}$ under the principal embedding. 

For a theory with non-full punctures, we can compute the character via the effect of Drinfeld-Sokolov reduction\cite{Beem:2014rza}, and in fact this just modifies the wavefunction. We can compute a variety of properties of the index via the first two terms. For 4d theories, if there are terms of order $\tau$ these indicate the presence of free hypermultiplets due to unitarity bounds, and at the VOA level, these correspond to spin $\frac{1}{2}$ $\beta \gamma$ systems. While we don't have such a justification for the 3d theories and VOAs we will be analyzing, we will assume that these class-S VOAs behave in the same way. In fact, in most cases we will be able to make a purely VOA justification of the above. Alternatively, one can compute the Hilbert series of the Coulomb branch of the corresponding quiver and use unitarity bounds for the monopole operators. 

For 4d theories, we can compute the flavour symmetry algebra of the theory via the order $\tau^2$ terms as those are moment map operators. From the VOA perspective these are just spin-1 currents. For the theories above, we can only see how they decompose in terms of the manifest symmetry of the fixture. The full flavour symmetry algebra has its adjoint decompose into the representations that show up at order $\tau^2$.

\subsection{\texorpdfstring{$C_2$}{C2} VOAs}

\subsubsection{Examples}
We begin our analysis with the simplest non-simply laced Lie algebra, that is $C_2$. We find five non-bad VOAs in total. The trinion has a symmetry $Sp(2)_6^3$ and 2d central charge $-56$. Doing a Drinfeld-Sokolov reduction of one of the $Sp(2)_6$ factors corresponding to the nilpotent orbit $[2^2]$ gives a VOA with central charge $-26$. Computing the Virasoro character we find 
\begin{equation}
    \begin{split}
        \mathcal{I}(\tau) = & 1+78 \tau ^2+2509 \tau ^4+49270 \tau ^6+698425 \tau ^8+7815106 \tau ^{10}+72903350 \tau ^{12}+  587906696 \tau ^{14} \\
         &+4204567965 \tau ^{16}+27174694560 \tau ^{18}+161016744070 \tau ^{20}+884547201850 \tau ^{22} \\ 
         &+4545922103619 \tau ^{24} +22017119036040 \tau ^{26}+O\left(\tau ^{27}\right). 
    \end{split}
\end{equation}

Thus, we expect that the symmetry was enhanced to an $E_6$ symmetry which can be confirmed by looking at the refined index at order $\tau^2$. We might suspect that this is just the $E_6$ current algebra at 2d level $-3$. Indeed comparing the character above with that of the current algebra we find they agree to at least order $\tau^{26}$.

Since our VOA has two full punctures, we can arbitrarily glue as many as we like together, and the resulting VOA should correspond to a perfectly good 4d SCFT. Indeed this same procedure is realized in the twisted $A_3$ theory \cite{Chacaltana:2012ch}, where the fixture with two twisted punctures and one simple puncture is also the $E_6$ rank-one instanton theory, and hence they can be glued together in that theory as well.

Let's look at other $C_2$ VOAs, starting with the trinion theory obtained by doing a Drinfeld-Sokolov reduction of one of the $Sp(2)_6$ factors corresponding to the nilpotent orbit $[2,1^2]$. This should have manifest flavour symmetry $Sp(2)_6^2\times SU(2)_5$. Computing its character we see that the flavour symmetry is enhanced to $Sp(4)_6 \times SU(2)_5$.  The central charge is $-39$. While this does not correspond to any known 4d $\mathcal{N}=2$ SCFT, we note that it saturates the 4d unitarity bound for the level of the $Sp(4)$ symmetry.

Doing a $[2,1^2]$ Drinfeld-Sokolov reduction of a manifest $Sp(2)_6$ in the above theory, we then obtain a VOA with manifest symmetry $Sp(2)_6\times SU(2)_5^2$. Note since the manifest symmetry of the parent VOA was enhanced, we expect the Drinfeld-Sokolov reduction to result in a VOA tensored with some spin $\frac{1}{2}$ beta gamma systems\cite{Distler:2022kjb}. Computing the character of the VOA and dividing by the contribution of the $\beta \gamma$ systems we get that the interacting part has an $F_4$ current algebra at 2d level $-\frac{5}{2}$. Computing the central charge after factoring out the $\beta \gamma$ systems we find $c_{2d} = -20$. Thus, we are led to believe that this may in fact be the rank-one $F_4$ instanton VOA. Computing the unrefined character to high order gives 
\begin{equation}
\begin{split}
    \mathcal{I}(\tau) = &1+52 \tau ^2+1106 \tau ^4+14808 \tau ^6+147239 \tau ^8+1183780 \tau ^{10}+8095998 \tau ^{12}+48688888 \tau ^{14} \\
     &+263508351 \tau^{16} +1305275544 \tau ^{18}+5993906570 \tau ^{20}+25771913376 \tau ^{22} \\ 
     &+104583612240 \tau ^{24}+403149160444 \tau ^{26}+O\left(\tau ^{27}\right)
\end{split}
\end{equation}
which agrees with the known character \cite{Beem:2017ooy}.

\subsubsection{Table}
We list the VOAs found in this section along with their various properties. 

\begin{longtable}{|c|c|c|c|c|}
\caption{$C_2$ Fixtures}\label{C2fixtures}\\
\hline
\#&Fixture& Flavour Symmetry&\begin{tabular}{c} Graded CB Dimensions \\ $\Delta_1,\Delta_2,...\Delta_r$\end{tabular}& ($n_h,n_v)$\\
\endfirsthead
\hline
\#&Fixture&Flavour Symmetry&\begin{tabular}{c} Graded CB Dimensions \\ $\Delta_1,\Delta_2,...\Delta_r$\end{tabular}& ($n_h,n_v)$\\
\endhead
\endfoot
\hline
1&$\begin{matrix} [1^4]\\ [1^4] \end{matrix}\quad [1^4]$&$\begin{gathered}{Sp(2)}_{6}^3\end{gathered}$&$\begin{gathered} \text{N/A} \end{gathered}$& (28,14)\\ 
\hline
2&$\begin{matrix} [2,1^2]\\ [1^4] \end{matrix}\quad [1^4]$&$\begin{gathered}{Sp(4)}_{6}\times SU(2)_5 \end{gathered}$&$\begin{gathered}\text{N/A} \end{gathered}$& (21,9)\\ 
\hline
3&$\begin{matrix} [2^2]\\ [1^4] \end{matrix}\quad [1^4]$&$\begin{gathered}(E_6)_6 \end{gathered}$&$\begin{gathered}3\end{gathered}$& (16,5)\\ 
\hline
4&$\begin{matrix} [2,1^2]\\ [2,1^2] \end{matrix}\quad [1^4]$&$\begin{gathered} (F_4)_5+2 \end{gathered}$&$\begin{gathered}\text{N/A}\end{gathered}$& (14,4)\\
\hline
5&$\begin{matrix} [1^4]\\ [2,1^2] \end{matrix}\quad [2^2]$&$\begin{gathered}(1,4)+\frac{1}{2}(2,5)\end{gathered}$&$\begin{gathered} \text{None} \end{gathered}$& (9,0)\\ 
\hline
\caption{$C_2$ VOAs and their would-be 4d invariants. For VOAs not corresponding to known 4d theories, we calculate their would-be anomalies from the change that occurs with nilpotent Higgsing. The plus indicates the number of decoupled $\beta\gamma$ systems. The central charge of the VOA is given by $-(2n_v+n_h)$. N/A in the Coulomb branch column indicates that a theory can be ruled out using the methods of Section \ref{Dualities}. None indicates the theory/VOA is just free hypermultiplets/ $\beta \gamma$ systems.}
\end{longtable}

\subsection{\texorpdfstring{$C_3$}{C3} VOAs}

\subsubsection{Examples}

In this case there are considerably more triples of nilpotent orbits that lead to non-bad VOAs. Thus, instead of analyzing all of them, we will only look at a small subset.

Consider the fixture with punctures $[1^6],[1^6]$ and $[4,2]$. The central charge is $-76$. Computing the Schur index we find the $Sp(3)_8^2$ symmetry is enhanced to $Sp(6)_8$. This matches with the $Sp(6)_8$ theory that appears in the twisted $D_4$ theory\cite{Chacaltana:2013oka}. We can do various Drinfeld-Sokolov reductions of one of the full punctures and obtain other theories matching those in the twisted $D_4$ sector.

Another example is the fixture $[1^6],[3^2],[2^3]$ which appears to be the VOA of the $E_7$ Minahan-Nemeschansky theory. We find a 2d central charge of $-38$ and check the characters agree to order $\tau^{20}$. The computed character is
\begin{equation}
    \begin{split}
        \mathcal{I}(\tau) = & 1+133 \tau ^2+7505 \tau ^4+254885 \tau ^6+6093490 \tau ^8+112077998 \tau ^{10} +1678245091 \tau ^{12} \\
        &+21264679635 \tau ^{14}+ 234433785700 \tau ^{16}+2296105563465 \tau ^{18} +20303111086038 \tau ^{20} \\
        &+O\left(\tau ^{21}\right).
    \end{split}
\end{equation}

From the above realization of $E_7$ we see that the fixture with punctures $[1^6],[2^3]$ and $[4,2]$ should be some free hypermultiplets. Let us consider going up to the fixture $[1^6],[2^3]$ and $[4,1^2]$. We find that the central charge agrees with the central charge of the $F_4$ rank-one instanton VOA tensored with 9 spin 1/2 beta gamma systems. Factoring their contribution from the character we find 
\begin{equation}
    1+52 \tau ^2+1106 \tau ^4+14808 \tau ^6+147239 \tau ^8+1183780 \tau ^{10}+O\left(\tau ^{11}\right)
\end{equation}
as expected. The VOA with punctures $[1^6],[4,1^2]$ and $[2^2,1^2]$ appears to be the $Sp(4)_6 \times SU(2)_5$ VOA seen in the $C_2$ case. Fixture 22 appears to be the $C_2$ type trinion.

\subsubsection{Table}

We list all VOAs corresponding to known 4d SCFTs and those with enhanced symmetries. We find more examples of theories that saturate the 4d unitarity bound for $Sp(n)$.

\begin{longtable}{|c|c|c|c|c|}
\caption{$C_3$ Fixtures}\label{C3fixtures}\\
\hline
\#&Fixture& Flavour Symmetry&\begin{tabular}{c} Graded CB Dimensions \\ $\Delta_1,\Delta_2,...\Delta_r$\end{tabular}& ($n_h,n_v)$\\
\endfirsthead
\hline
\#&Fixture& Flavour Symmetry&\begin{tabular}{c} Graded CB Dimensions \\ $\Delta_1,\Delta_2,...\Delta_r$\end{tabular}& $(n_h,n_v)$\\
\endhead
\endfoot
\hline
1&$\begin{matrix} [1^6]\\ [1^6] \end{matrix}\quad [3^2]$&$\begin{gathered}{Sp(3)}_{8}^2 \times SU(2)_8\end{gathered}$&$\begin{gathered}4,4,6\end{gathered}$& (48,25)\\ 
\hline
2&$\begin{matrix} [1^6]\\ [2,1^4] \end{matrix}\quad [3^2]$&$\begin{gathered}{Sp(4)}_{8} \times Sp(2)_7\end{gathered}$&$\begin{gathered}4,6\end{gathered}$& (38,18)\\ 
\hline
3&$\begin{matrix} [1^6]\\ [1^6] \end{matrix}\quad [4,2]$&$\begin{gathered}{Sp(6)}_{8}\end{gathered}$&$\begin{gathered}4,6\end{gathered}$& (40,18)\\ 
\hline
4&$\begin{matrix} [1^6]\\ [2,1^4] \end{matrix}\quad [4,2]$&$\begin{gathered}{Sp(5)}_{7} +3 \end{gathered}$&$\begin{gathered}6 \end{gathered}$& (30,11)\\ 
\hline
5&$\begin{matrix} [1^6]\\ [2^2,1^2] \end{matrix}\quad [4,2]$&$\begin{gathered}{(E_6)}_{6}+6 \end{gathered}$&$\begin{gathered} 3 \end{gathered}$& (22,5)\\ 
\hline
6&$\begin{matrix} [2^3]\\ [1^6] \end{matrix}\quad [4,2]$&$\begin{gathered}\frac{1}{2}(1,14')+\frac{1}{2}(3,6) \end{gathered}$&$\begin{gathered}\text{None}\end{gathered}$& (16,0)\\
\hline
7&$\begin{matrix} [2^3]\\ [3^2] \end{matrix}\quad [1^6]$&$\begin{gathered}(E_7)_8 \end{gathered}$&$\begin{gathered} 4 \end{gathered}$& (24,7)\\ 
\hline
8&$\begin{matrix} [2^3]\\ [3^2] \end{matrix}\quad [2,1^4]$&$\begin{gathered}\frac{1}{2}(3,4,1)+\frac{1}{2}(1,5,2)+\frac{1}{2}(3,1,2) \end{gathered}$&$\begin{gathered} \text{None} \end{gathered}$& (14,0)\\ 
\hline
9&$\begin{matrix} [2^2,1^2]\\ [3^2] \end{matrix}\quad [1^6]$&$\begin{gathered}SU(8)_8 \times SU(2)_6 \end{gathered}$&$\begin{gathered}3,4\end{gathered}$& (30,12)\\ 
\hline
10&$\begin{matrix} [2^2,1^2]\\ [3^2] \end{matrix}\quad [2,1^4]$&$\begin{gathered}(E_6)_6+4 \end{gathered}$&$\begin{gathered} 3 \end{gathered}$& (20,5)\\ 
\hline
11&$\begin{matrix} [2^3]\\ [4,1^2] \end{matrix}\quad [1^6]$&$\begin{gathered}(F_4)_5+9 \end{gathered}$&$\begin{gathered} \text{N/A} \end{gathered}$& (21,4)\\
\hline
12&$\begin{matrix} [2^2,1^2]\\ [4,1^2] \end{matrix}\quad [1^6]$&$\begin{gathered}Sp(4)_6 \times SU(2)_5 +6 \end{gathered}$&$\begin{gathered} \text{N/A} \end{gathered}$& (27,9)\\
\hline
13&$\begin{matrix} [2,1^4]\\ [4,1^2] \end{matrix}\quad [1^6]$&$\begin{gathered}Sp(5)_7 \times SU(2)_5 +3 \end{gathered}$&$\begin{gathered} \text{N/A} \end{gathered}$& (35,15)\\
\hline
14&$\begin{matrix} [1^6]\\ [4,1^2] \end{matrix}\quad [1^6]$&$\begin{gathered}Sp(6)_8 \times SU(2)_5  \end{gathered}$&$\begin{gathered} \text{N/A} \end{gathered}$& (45,22)\\
\hline
15&$\begin{matrix} [2,1^4]\\ [2,1^4] \end{matrix}\quad [2,1^4]$&$\begin{gathered}Sp(2)_{7}^{3} \times U(1)  \end{gathered}$&$\begin{gathered} \text{N/A} \end{gathered}$& (58,37)\\
\hline
16&$\begin{matrix} [2^2,1^2]\\ [2,1^4] \end{matrix}\quad [2,1^4]$&$\begin{gathered}Sp(2)_{7}^{2} \times SU(2)_6 \times SU(2)_k  \end{gathered}$&$\begin{gathered} \text{N/A} \end{gathered}$& (50,31)\\
\hline
17&$\begin{matrix} [2^2,1^2]\\ [2,1^4] \end{matrix}\quad [2^2,1^2]$&$\begin{gathered}Sp(2)_{7} \times SU(2)_6^2 \times SU(2)_{12}^2 \end{gathered}$&$\begin{gathered} \text{N/A} \end{gathered}$& (42,25)\\
\hline
18&$\begin{matrix} [2^2,1^2]\\ [2^2,1^2] \end{matrix}\quad [2^2,1^2]$&$\begin{gathered} SU(2)_6^7 \end{gathered}$&$\begin{gathered} \text{N/A} \end{gathered}$& (34,19)\\
\hline
19&$\begin{matrix} [2^3]\\ [2,1^4] \end{matrix}\quad [2,1^4]$&$\begin{gathered}Sp(2)_{7}^2 \times SU(2)_{12}^2 \end{gathered}$&$\begin{gathered} \text{N/A} \end{gathered}$& (44,26)\\
\hline
20&$\begin{matrix} [2^3]\\ [2^3] \end{matrix}\quad [2,1^4]$&$\begin{gathered}SU(4)_{12} \times SU(4)_7  \end{gathered}$&$\begin{gathered} \text{N/A} \end{gathered}$& (30,15)\\
\hline
21&$\begin{matrix} [2^3]\\ [2^3] \end{matrix}\quad [2^2,1^2]$&$\begin{gathered}Sp(4)_{6} \times SU(2)_5 +1 \end{gathered}$&$\begin{gathered} \text{N/A} \end{gathered}$& (22,9)\\
\hline
22&$\begin{matrix} [2^3]\\ [2,1^4] \end{matrix}\quad [2^2,1^2]$&$\begin{gathered}Sp(2)_{12} \times Sp(2)_7 \times SU(2)_6 \end{gathered}$&$\begin{gathered} \text{N/A} \end{gathered}$& (36,20)\\
\hline
23&$\begin{matrix} [2^3]\\ [2^2,1^2] \end{matrix}\quad [2^2,1^2]$&$\begin{gathered}Sp(2)_6^3 \end{gathered}$&$\begin{gathered} \text{N/A} \end{gathered}$& (28,14)\\
\hline
\caption{$C_3$ VOAs and their would-be 4d invariants. The plus indicates the number of decoupled $\beta\gamma$ systems. The central charge of the VOA is given by $-(2n_v+n_h)$. N/A in the Coulomb branch column indicates that a theory can be ruled out using the methods of Section \ref{Dualities}.}
\end{longtable}

\subsection{\texorpdfstring{$G_2$}{G2} VOAs}

We now turn our attention to the $G_2$ VOAs, here we list the three punctured spheres with enhanced symmetries as well as previously known VOAs. In Table \ref{G2} fixtures 1-6 appear in the $\mathbb{Z}_3$ twisted $D_4$ theory \cite{Chacaltana:2016shw}  while fixtures 3-5, 7 and 8 appear in the $S_3$ twisted $D_4$ theory\cite{Distler:2021cwz}. In the $D_4$ theory with non-abelian twists, fixtures are labeled by two nilpotent orbits of $C_3$ and one of $G_2$. Additionally, the quivers attached to the central node for the $[4,1^2]$ puncture and the $\tilde{A}_1$ puncture are the same, which we suspect is related to the overlap of the VOAs found here and in the $S_3$ twisted case.

\subsubsection{Table}

\begin{longtable}{|c|c|c|c|c|}
\caption{$G_2$ Fixtures}\label{G2fixtures}\\
\hline
\#&Fixture& Flavour Symmetry&\begin{tabular}{c} Graded CB Dimensions \\ $\Delta_1,\Delta_2,...\Delta_r$\end{tabular}& ($n_h,n_v)$\\
\endfirsthead
\hline
\#&Fixture& Flavour Symmetry&\begin{tabular}{c} Graded CB Dimensions \\ $\Delta_1,\Delta_2,...\Delta_r$\end{tabular}& $(n_h,n_v)$\\
\endhead
\endfoot
\hline
1&$\begin{matrix} 0\\ 0 \end{matrix}\quad G_2(a_1)$&$\begin{gathered}(G_2)_8 \times (G_2)_8\end{gathered}$&$\begin{gathered}4,4,6\end{gathered}$& (40,25)\\ 
\hline
2&$\begin{matrix} 0\\ A_1 \end{matrix}\quad G_2(a_1)$&$\begin{gathered}(G_2)_8 \times SU(2)_{14}\end{gathered}$&$\begin{gathered}4,6\end{gathered}$& (30,18)\\ 
\hline
3&$\begin{matrix} 0\\ \tilde{A_1} \end{matrix}\quad G_2(a_1)$&$\begin{gathered}Spin(8)_8 \times SU(2)_{5}\end{gathered}$&$\begin{gathered}4,2\end{gathered}$& (21,10)\\ 
\hline
4&$\begin{matrix} 0\\ G_2(a_1) \end{matrix}\quad G_2(a_1)$&$\begin{gathered}Spin(8)_4 \times Spin(8)_4 \end{gathered}$&$\begin{gathered}2,2\end{gathered}$& (16,6)\\ 
\hline
5&$\begin{matrix} \tilde{A_1} \\ A_1 \end{matrix}\quad G_2(a_1)$&$\begin{gathered}Spin(8)_4 +3\end{gathered}$&$\begin{gathered}2\end{gathered}$& (11,3)\\ 
\hline
6&$\begin{matrix} A_1 \\ A_1 \end{matrix}\quad G_2(a_1)$&$\begin{gathered}SU(4)_{14}\end{gathered}$&$\begin{gathered}6\end{gathered}$& (20,11)\\ 
\hline
7&$\begin{matrix} \tilde{A_1} \\ \tilde{A_1} \end{matrix}\quad 0$&$\begin{gathered}Spin(7)_8 \times SU(2)_5^2\end{gathered}$&$\begin{gathered}4,4\end{gathered}$& (26,14)\\ 
\hline
8&$\begin{matrix} \tilde{A_1} \\ \tilde{A_1} \end{matrix}\quad A_1$&$\begin{gathered}SU(2)_8 \times Sp(3)_5+1\end{gathered}$&$\begin{gathered}4\end{gathered}$& (16,7)\\ 
\hline
9&$\begin{matrix} \tilde{A_1} \\ A_1 \end{matrix}\quad A_1$&$\begin{gathered}Sp(2)_{14} \times SU(2)_5\end{gathered}$&$\begin{gathered}\text{N/A}\end{gathered}$& (25,15)\\ 
\hline
\caption{$G_2$ fixtures corresponding to known 4d theories and those with enhanced symmetries. The plus indicates the number of decoupled $\beta\gamma$ systems. N/A in the Coulomb branch column indicates that a theory can be ruled out using the methods of Section \ref{Dualities}.}
\label{G2}
\end{longtable}

\subsection{\texorpdfstring{$B_3$}{B3} VOAs}

\subsubsection{Examples}

Let us now turn our attention to the $B_3$ VOAs. The fixture with two full punctures and a simple puncture has central charge -64 and flavour symmetry $Spin(14)_{10}$ which agrees with the corresponding twisted $A_5$ fixture.

Now consider the theory with punctures $[1^7],[3^2,1]$ and $[3,2^2]$. It has manifest symmetry $Spin(7)_{10} \times SU(2)_6 \times U(1)$ that is enhanced to $(F_4)_{10}\times SU(2)_6$. It also has central charge $-56$. This matches precisely with the data of the rank-two $F_4$ instanton VOA \cite{Beem:2019snk}. There they compute the character up to order $\tau^7$ and in \cite{Gu:2019dan} it is computed to order $\tau^{11}$. We find perfect agreement comparing these computations to ours given by 

\begin{equation}
\begin{split}
   \mathcal{I}(\tau) = &1+55 \tau ^2+104 \tau ^3+1595 \tau ^4+5072 \tau ^5+35226 \tau ^6+130240 \tau ^7+640886 \tau ^8+2384608 \tau ^9 \\
   & +9769738 \tau ^{10}+ 34831256 \tau ^{11}+127101065 \tau ^{12}+428834560 \tau ^{13}+1439899326 \tau ^{14} \\ 
   &+4598638800 \tau ^{15}+14466877609 \tau ^{16}+O\left(\tau ^{17}\right). 
\end{split}
\end{equation}

In certain cases, we can perform a nilpotent Higgsing/Drinfeld-Sokolov reduction of the symmetry of a non-full puncture \cite{Distler:2022nsn,Distler:2022kjb,DistlerElliotWIP} which has the effect of replacing it with another puncture. In this case, doing a Drinfeld-Sokolov reduction of the $SU(2)_6$ replaces the $[3,2^2]$ puncture with $[3^2,1]$, and appears to give the product of two rank-one $F_4$ instanton VOAs. Additionally, we could do a Drinfeld-Sokolov reduction of the manifest $Spin(7)_{10}$ to obtain a rank-one $F_4$ instanton VOA plus some $\beta \gamma$ systems. Thus, we see the expected behavior under Drinfeld-Sokolov reduction. 
\subsubsection{Table}
We list all VOAs with enhanced symmetries in the table below. There appears to be substantial overlap with the $C_3$ case. Fixtures 5 and 6 appear to be isomorphic. 

\begin{longtable}{|c|c|c|c|c|}
\caption{$B_3$ Fixtures}\label{B3fixtures}\\
\hline
\#&Fixture& Flavour Symmetry&\begin{tabular}{c} Graded CB Dimensions \\ $\Delta_1,\Delta_2,...\Delta_r$\end{tabular}& ($n_h,n_v)$\\
\endfirsthead
\hline
\#&Fixture& Flavour Symmetry&\begin{tabular}{c} Graded CB Dimensions \\ $\Delta_1,\Delta_2,...\Delta_r$\end{tabular}& $(n_h,n_v)$\\
\endhead
\endfoot
\hline
1&$\begin{matrix} [1^7]\\ [1^7] \end{matrix}\quad [5,1^2]$&$\begin{gathered}{Spin(14)}_{10} \times U(1)\end{gathered}$&$\begin{gathered}3,5\end{gathered}$& (36,14)\\ 
\hline
2&$\begin{matrix} [1^7]\\ [2^2,1^3] \end{matrix}\quad [5,1^2]$&$\begin{gathered}(E_6)_6+7\end{gathered}$&$\begin{gathered}3\end{gathered}$& (23,5)\\ 
\hline
3&$\begin{matrix} [1^7]\\ [3,1^4] \end{matrix}\quad [3,1^4]$&$\begin{gathered}{Spin(8)}_{10} \times SU(2)_6^4 \end{gathered}$&$\begin{gathered}\text{N/A}\end{gathered}$& (48,28)\\ 
\hline
4&$\begin{matrix} [1^7]\\ [3,2^2] \end{matrix}\quad [3,1^4]$&$\begin{gathered}{Spin(8)}_{10} \times SU(2)_6^3 \end{gathered}$&$\begin{gathered}\text{N/A}\end{gathered}$& (42,23)\\ 
\hline
5&$\begin{matrix} [1^7]\\ [3,2^2] \end{matrix}\quad [3,2^2]$&$\begin{gathered}{Spin(9)}_{10} \times SU(2)_6^2 \end{gathered}$&$\begin{gathered}\text{N/A}\end{gathered}$& (36,18)\\ 
\hline
6&$\begin{matrix} [1^7]\\ [3^2,1] \end{matrix}\quad [3,1^4]$&$\begin{gathered}{Spin(9)}_{10} \times SU(2)_6^2 \end{gathered}$&$\begin{gathered}\text{N/A}\end{gathered}$& (36,18)\\ 
\hline
7&$\begin{matrix} [1^7]\\ [3^2,1] \end{matrix}\quad [3,2^2]$&$\begin{gathered}(F_4)_{10} \times SU(2)_6 \end{gathered}$&$\begin{gathered}\text{N/A}\end{gathered}$& (30,13)\\ 
\hline
8&$\begin{matrix} [1^7]\\ [3^2,1] \end{matrix}\quad [3^2,1]$&$\begin{gathered}(F_4)_{5}^2  \end{gathered}$&$\begin{gathered}\text{N/A}\end{gathered}$& (24,8)\\ 
\hline
9&$\begin{matrix} [2^2,1^3]\\ [3^2,1] \end{matrix}\quad [3,2^2]$&$\begin{gathered}(F_4)_{5}+5 \end{gathered}$&$\begin{gathered}\text{N/A}\end{gathered}$& (17,4)\\ 
\hline
10&$\begin{matrix} [2^2,1^3]\\ [2^2,1^3] \end{matrix}\quad [3,1^4]$&$\begin{gathered}Sp(2)_{7} \times SU(2)_{12}^2 \times SU(2)_6^2  \end{gathered}$&$\begin{gathered}\text{N/A}\end{gathered}$& (42,25)\\ 
\hline
11&$\begin{matrix} [2^2,1^3]\\ [2^2,1^3] \end{matrix}\quad [3,2^2]$&$\begin{gathered}Sp(2)_{12} \times Sp(2)_7 \times SU(2)_6\end{gathered}$&$\begin{gathered}\text{N/A}\end{gathered}$& (36,20)\\ 
\hline
12&$\begin{matrix} [2^2,1^3]\\ [2^2,1^3] \end{matrix}\quad [3^2,1]$&$\begin{gathered}SU(4)_{12} \times SU(4)_7 \end{gathered}$&$\begin{gathered}\text{N/A}\end{gathered}$& (30,15)\\ 
\hline
13&$\begin{matrix} [2^2,1^3]\\ [3,1^4] \end{matrix}\quad [3,1^4]$&$\begin{gathered} SU(2)_6^7+1\end{gathered}$&$\begin{gathered}\text{N/A}\end{gathered}$& (35,19)\\ 
\hline
14&$\begin{matrix} [2^2,1^3]\\ [3,2^2] \end{matrix}\quad [3,1^4]$&$\begin{gathered}Sp(2)_6^3+1\end{gathered}$&$\begin{gathered}\text{N/A}\end{gathered}$& (29,14)\\ 
\hline
15&$\begin{matrix} [2^2,1^3]\\ [3^2,1] \end{matrix}\quad [3,1^4]$&$\begin{gathered}Sp(4)_6 \times SU(2)_5+2\end{gathered}$&$\begin{gathered}\text{N/A}\end{gathered}$& (23,9)\\ 
\hline
16&$\begin{matrix} [2^2,1^3]\\ [3,2^2] \end{matrix}\quad [3,2^2]$&$\begin{gathered}Sp(4)_6 \times SU(2)_5+2\end{gathered}$&$\begin{gathered}\text{N/A}\end{gathered}$& (23,9)\\ 
\hline
\caption{$B_3$ VOAs and their would-be 4d invariants. The plus indicates the number of decoupled $\beta\gamma$ systems. The central charge of the VOA is given by $-(2n_v+n_h)$. N/A in the Coulomb branch column indicates that a theory can be ruled out using the methods of Section \ref{Dualities}.}
\end{longtable}

\subsection{Higher Rank Outlook}
One could continue analyzing these VOAs ad nauseam, though the higher the rank of the Lie algebra, the more difficult it is to compute the characters. In light of our conjecture on the overlap with twisted class-S VOAs, we check that it is consistent at the level of central charges. For example, for $F_4$ we find the VOA with two full punctures and a subregular orbit puncture has $c_{2d}=-398$ and flavour symmetry $(F_4)_{18}^2$, which agrees with the corresponding twisted $E_6$ theory\cite{Chacaltana:2015bna}.

For $B_n$ and $C_n$, one can perform a similar computation and find the fixtures with two full punctures and a subregular orbit puncture have 2d central charges $-8n^2+2n+2$ and $26-10n-8n^2$ respectively. Additionally, we see that the flavour symmetries are enhanced to $Spin(4n+2)_{4n-2}\times U(1)$ and $Sp(2n)_{2n+2}$ respectively. In the $B_n$ case this is just the $R_{2,2n-1}$ theory.

We note there appears to be an infinite family of $Sp(2n)_{2n+2} \times SU(2)_5$ theories obtained from the $C_n$ fixtures with two full punctures and a $[2n-2,1^2]$ puncture. These all saturate a 4d universal lower bound on the level of the $Sp(2n)$ symmetry. Performing a Drinfeld-Sokolov reduction of the manifest $Sp(n)$ then gives some $\beta \gamma$ systems tensored with a $Sp(2n+1)_{2n+3}\times SU(2)_5$ VOA.

Unfortunately, it's not obvious if there are higher rank $F_4$ instanton VOAs amongst the non-simply laced class-S VOAs. Nevertheless, one can find more realizations of the $F_4$ instanton VOAs we have seen so far. In the $B_4$ case for example, there are six realizations of $F_4$ instanton VOAs.

\begin{longtable}{|c|c|c|c|c|}
\caption{$B_4$ VOAs}\label{B4fixtures}\\
\hline
\#&Fixture& Manifest Flavour Symmetry&\begin{tabular}{c} Flavour Symmetry \end{tabular}& ($n_h,n_v)$\\
\endfirsthead
\hline
\#&Fixture& Manifest Flavour Symmetry&\begin{tabular}{c} Flavour Symmetry \end{tabular}& ($n_h,n_v)$\\
\endhead
\endfoot
\hline
1&$\begin{matrix} [1^9]\\ [4^2,1] \end{matrix}\quad [5,2^2]$&$\begin{gathered}Spin(9)_{14} \times SU(2)_9 \times SU(2)_6\end{gathered}$&$\begin{gathered} (F_4)_{10}\times SU(2)_6+\frac{1}{2}(9,2,1) \end{gathered}$& (39,13)\\ 
\hline
2&$\begin{matrix} [3,1^6]\\ [3^3] \end{matrix}\quad [4^2,1]$&$\begin{gathered}Spin(6)_{10} \times SU(2)_{28} \times SU(2)_9 \end{gathered}$&$\begin{gathered}(F_4)_{10}\times SU(2)_6+\frac{1}{2}(1,3,2) \end{gathered}$& (33,13)\\ 
\hline
3&$\begin{matrix} [3,2^2,1^2]\\ [3^3] \end{matrix}\quad [4^2,1]$&$\begin{gathered}SU(2)_8 \times U(1) \times SU(2)_{28} \times SU(2)_9 \end{gathered}$&$\begin{gathered}(F_4)_{5}+9 \end{gathered}$& (21,4)\\ 
\hline
4&$\begin{matrix} [2^2,1^5]\\ [3^3] \end{matrix}\quad [5,3,1]$&$\begin{gathered}Spin(5)_{10} \times SU(2)_{28} \times SU(2)_9 \end{gathered}$&$\begin{gathered}(F_4)_{10}\times SU(2)_6+\frac{1}{2}(1,3,2)\end{gathered}$& (33,13)\\ 
\hline
5&$\begin{matrix} [2^4,1]\\ [3^3] \end{matrix}\quad [5,3,1]$&$\begin{gathered}Sp(2)_{9} \times SU(2)_{28} \times SU(2)_9 \end{gathered}$&\begin{tabular}{c}$\begin{gathered}(F_4)_{5}+10\end{gathered}$\end{tabular}& (22,4)\\ 
\hline
6&$\begin{matrix} [1^9]\\ [4^2,1] \end{matrix}\quad [5,3,1]$&$\begin{gathered}Spin(9)_{14} \times SU(2)_9\end{gathered}$&$\begin{gathered} (F_4)_{5}^2+\frac{1}{2}(9,2) \end{gathered}$& (33,8)\\ 
\hline
\caption{$F_4$ Instanton VOAs found in the $B_4$ theory. We don't list the representations of the hypermultiplets in fixtures three and five.}
\end{longtable}
While we leave a complete investigation into the $F_4$ VOAs for future work, we point out the following two instances that appear to contain products of $F_4$ instanton VOAs with an $E_7$ instanton VOA. 
\begin{longtable}{|c|c|c|c|c|}
\caption{$F_4$ VOAs}\label{F4fixtures}\\
\hline
\#&Fixture& Manifest Flavour Symmetry&\begin{tabular}{c} Flavour Symmetry \end{tabular}& ($n_h,n_v)$\\
\endfirsthead
\hline
\#&Fixture& Manifest Flavour Symmetry&\begin{tabular}{c} Flavour Symmetry \end{tabular}& ($n_h,n_v)$\\
\endhead
\endfoot
\hline
1&$\begin{matrix} 0\\ C_3 \end{matrix}\quad B_3$&$\begin{gathered}(F_4)_{18} \times SU(2)_6 \times SU(2)_{24} \end{gathered}$&$\begin{gathered} (F_4)_{10}\times SU(2)_6 \times (E_7)_8 \end{gathered}$& (39,20)\\ 
\hline
2&$\begin{matrix} 0\\ F_4(a_2) \end{matrix}\quad B_3$&$\begin{gathered}(F_4)_{18} \times SU(2)_{24}  \end{gathered}$&$\begin{gathered}(F_4)_{5}^2\times (E_7)_8 \end{gathered}$& (33,15)\\ 
\hline
\caption{Products of $F_4$ instanton VOAs with the rank-one $E_7$ instanton VOA in the $F_4$ VOAs.}
\end{longtable}

\section{Quiver Gauge Theories}\label{Quivers}

So far we have only focused on the VOAs themselves. However, from the proposal of \cite{Beem:2022mde}, these should correspond to certain quiver gauge theories, which are interesting in their own right. We expect the associated varieties of these VOAs to be the Coulomb branches of the 3d theories. 

One might ask whether other objects might carry over from the physical class-S perspective. We expect the Hilbert series formula to carry over straightforwardly for genus zero theories, that is we just use the wavefunctions that appear in twisted class-S theories. For a genus zero theory with $n$ punctures the Hilbert series is given by

\begin{equation}
    \mathcal{I}_{\text{Hall-Littlewood}} = \sum_{\mathfrak{R}}\frac{\product_{i=1}^n\psi^{\rho_i}_{\mathfrak{R}}(\mathbf{a}_i,\tau)}{\big(\psi^{\rho}_{\mathfrak{R}}(\tau)\big)^{n-2}}
\end{equation}
where the sum is over irreducible representations of the corresponding Lie algebra and we use the twisted Hall-Littlewood wavefunctions\cite{Mekareeya:2012tn}. Here $\rho_i$ denotes the corresponding homomorphism for the $i$th puncture and $\rho$ denotes the principal embedding. This can be obtained from the gluing procedure in \cite{Cremonesi:2014vla}.

\begin{figure}
\centering
\begin{tikzpicture}
\node (0) at (0,0) {$O(2)$};
\node (A1) at (2.5,0) {$Sp(1)$};
\node (tA1) at (5,0) {$O(4)$};
\node (G2a1) at (7.5,0) {$Sp(2)$};
\node (G2) at (10,-1) {$O(4)$};
\node (a) at (10,1) {$O(4)$};
\path[thick,draw,color=black] 
(0) edge  (A1)
(A1) edge  (tA1)
(tA1) edge (G2a1)
(G2a1) edge  (G2)
(G2a1) edge (a)
;
\end{tikzpicture}
\caption{Quiver gauge theory whose Coulomb branch is the product of $\mathbb{H}^2$ with the reduced moduli space of a single $F_4$ instanton.}
\label{fig:1F4}
\end{figure}
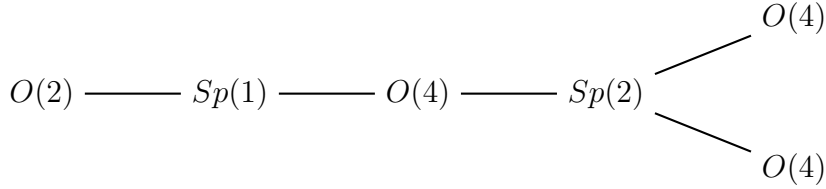

Applying this\footnote{We use Lie Art\cite{Feger:2019tvk} and the Mathematica package found in \cite{Ergun:2021wok}.} to the $C_2$ realization of the one $F_4$ instanton VOA and removing the contribution of the free hypers we find

\begin{equation}
    1+52 \tau ^2+1053 \tau ^4+12376 \tau ^6+100776 \tau ^8+627912 \tau ^{10}+3187041 \tau ^{12}+O\left(\tau ^{14}\right)
\end{equation}
in perfect agreement with known expressions. This is just the Hilbert series of the Coulomb branch of the star shaped quivers with two $T_{[2^2,1]}(SO(5))$ quivers and one $T_{[1^5]}(SO(5))$ quiver glued to the central $Sp(2)$ node as shown in Figure \ref{fig:1F4}. We can compute the dimension of the Higgs branch of this quiver, which should be equal to the complex dimension of the Coulomb branch of the parent 4d theory of the mirror 3d theory, if such a parent theory existed. We find the dimension is zero, suggesting that if it existed, the 4d theory with the $F_4$ instanton VOA is rank-zero. It should be noted that rank-zero theories are thought not to exist\cite{Argyres:2020nrr,Argyres:2015gha}, and hence this can be taken as additional evidence that there are no 4d $F_4$ instanton theories. We will see a more persuasive argument in the next section.

\begin{figure}
\centering
\begin{tikzpicture}
\node (0) at (0,0) {$O(2)$};
\node (A1) at (2,0) {$Sp(1)$};
\node (tA1) at (4,0) {$O(4)$};
\node (G2a1) at (6,0) {$Sp(2)$};
\node (G2) at (8,0) {$O(6)$};
\node (a) at (10,0) {$Sp(3)$};
\node (b) at (12,0) {$O(4)$};
\node (c) at (14,0) {$Sp(1)$};
\node (d) at (10,2) {$O(4)$};
\path[thick,draw,color=black] 
(0) edge  (A1)
(A1) edge  (tA1)
(tA1) edge (G2a1)
(G2a1) edge  (G2)
(G2) edge (a)
(a) edge (b) 
(b) edge (c)
(a) edge (d)
;
\end{tikzpicture}
\caption{Quiver gauge theory whose Coulomb branch is the reduced moduli space of two $F_4$ instantons.}
\label{2F4}
\end{figure}
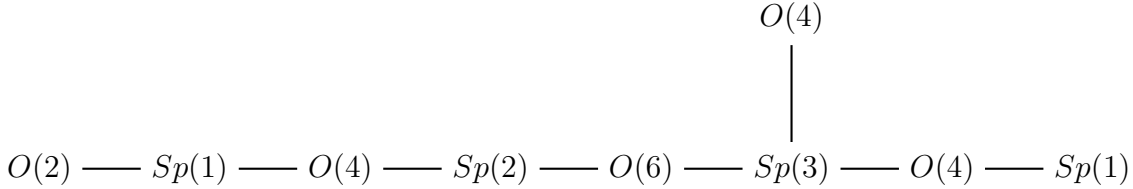

Using the $B_3$ realization of the two $F_4$ instanton theory we find the following Hilbert series of the Coulomb branch
\begin{equation}
 1+55 \tau ^2+104 \tau ^3+1539 \tau ^4+4966 \tau ^5+32091 \tau ^6+119340 \tau ^7+O\left(\tau ^8\right)
\end{equation}
which agrees with \cite{Hanany:2012dm}(See also \cite{Keller:2012da}). The corresponding 3d quiver gauge theory is shown in figure \ref{2F4} and has the shape of an extended $E_8$ Dynkin diagram. We can compute the dimension of the Higgs branch of this theory, and we find it to be one-dimensional. The Higgs branch of the quiver has Hilbert series\footnote{The Higgs branch of the quiver is a hyperkh\"{a}ler quotient of the corresponding nilpotent orbits and its Hilbert series can be computed using the results of \cite{Hanany:2016gbz}.}
\begin{equation}
    1+\tau^{12}+\tau^{20}+\tau^{24}+\tau^{30}+O(\tau^{32})
\end{equation}
which suggests it is just the $E_8$ surface singularity.

\begin{figure}
\centering
\begin{tikzpicture}
\node (0) at (0,0) {$O(2)$};
\node (A1) at (2.5,0) {$Sp(1)$};
\node (tA1) at (5,0) {$O(4)$};
\node (G2a1) at (7.5,0) {$Sp(2)$};
\node (G2) at (10,0) {$O(4)$};
\node (a) at (7.5,2) {$O(4)$};
\node (b) at (12.5,0) {$Sp(1)$};
\node (c) at (15,0) {$O(2)$};
\path[thick,draw,color=black] 
(0) edge  (A1)
(A1) edge  (tA1)
(tA1) edge (G2a1)
(G2a1) edge  (G2)
(G2a1) edge (a)
(G2) edge (b)
(b) edge (c)
;
\end{tikzpicture}
\caption{Quiver gauge theory for the $Sp(4)_6\times SU(2)_5$ VOA.}
\label{Sp(2)}
\end{figure}
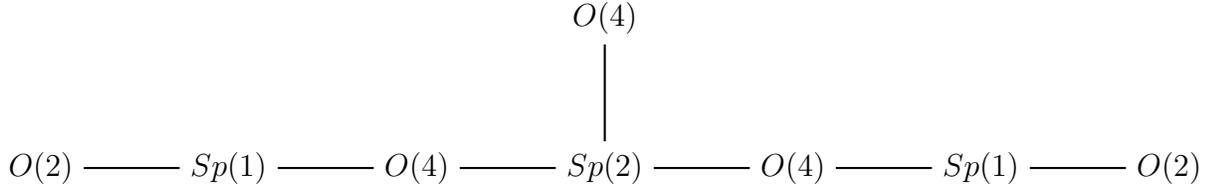

For the $Sp(4)_6\times SU(2)_5$ theory the quiver has the shape of an extended $E_7$ Dynkin diagram as shown in Figure \ref{Sp(2)}, and computing the Hilbert series of the Higgs branch, we find what appears to be the $E_7$ surface singularity. 

\label{Sp2}

\section{Dualities}\label{Dualities}

\subsection{Gluing Full Punctures}

In four dimensions, performing an exactly marginal gauging corresponds to gauging the current algebra of the VOA after tensoring with some $bc$ ghosts \cite{Beem:2013sza}. The condition for the 4d gauging to be exactly marginal corresponds to the level of the diagonal current algebra vanishing. One might then wonder what the interpretation of such gaugings are for VOAs considered here. Presumably these correspond to the VOA\footnote{As the exact interpretation of these VOAs is unclear, the notion of associating VOAs to a 3d SCFT is a bit awkward. Still we suspect in all these instances that the free field realizations of \cite{Beem:2019tfp} applied to the Coulomb branch of the 3d IR theory give the VOA obtained from gauging.} of the SCFT obtained by performing the gauging and flowing to the IR. The dualities seen at the VOA level should then correspond to IR dualities. We could also consider the same BRST reductions involving VOAs of known 4d theories and these non-simply laced ones. For the gaugings of symmetries from full punctures, such dualities are easily understood, at least at the level of characters. As an example, let's consider taking the product of the trinion of the $C_2$ VOAs with the class-S VOA corresponding to a hypermultiplet in the vector of $Spin(6)$ and the fundamental of $Sp(2)$. This corresponds to the fixture in the twisted $D_3$ theory with an untwisted full, a twisted full, and a twisted regular puncture. We will refer to this latter VOA in the twisted $D_3$ case and other class-S types as the hybrid cylinder. We could consider gauging the diagonal $Sp(2)_{12}$ symmetry and ask, what is the resulting VOA? Letting $\tilde{\psi}_{\mathfrak{R}}$ denote type $\mathfrak{j}$ wavefunctions and $\psi_{\mathfrak{R}}$ denote type $\mathfrak{g}$ wavefunctions, the resulting character is 
\begin{equation}
   \int dG \times \mathcal{I}_{vect} \sum_{\mathfrak{R}}\frac{(\psi^{[1^4]})^2\psi^{[1^4]}_{\mathfrak{R}}}{\psi^{[4]}_{\mathfrak{R}}} \times \sum_{\mathfrak{R}'}\frac{\psi^{[1^4]}_{\mathfrak{R}'}\tilde{\psi}^{[1^6]}_{\mathfrak{R}'}\psi^{[4]}_{\mathfrak{R}'}}{\tilde{\psi}^{[5,1]}_{\mathfrak{R}'}} =  \sum_{\mathfrak{R}}\frac{(\psi^{[1^4]})^2\tilde{\psi}^{[1^6]}_{\mathfrak{R}}}{\tilde{\psi}^{[5,1]}_{\mathfrak{R}}}
\end{equation}
where we integrate over the Haar measure and have used 
\begin{equation}
    \int dG \times \mathcal{I}_{vect} \times \psi^{\rho}_{\mathfrak{R}}(\mathbf{a},\tau) \psi^{\rho'}_{\mathfrak{R'}}(\mathbf{a},\tau)=\delta_{\mathfrak{R}\mathfrak{R}'}.
\end{equation}
On the right hand side we now have the index of the twisted $D_3 \cong A_3$ trinion. Thus, we see the VOA of the twisted trinion seems to be the BRST gauging of two other VOAs, and in the next subsection we will see they are in fact the same. This is somewhat surprising because the four-dimensional theory is an isolated SCFT. Hence, we have another indication that the $C_2$ trinion does not correspond to a 4d SCFT, as if it did, the above gauging should be possible, and would result in a theory with a non-trivial conformal manifold. This argument obviously generalizes to any of the non-simply laced trinions and shows how they do not correspond to 4d $\mathcal{N}=2$ SCFTs. Additionally one can make the same argument for theories with reduced punctures if they have at least one full puncture. For example, coupling the free hypermultiplets to the $F_4$ instanton theory and gauging the $Sp(2)_{12}$ results in the VOA of the rank two $Spin(7)_8 \times SU(2)_5^2$ theory, which is also an isolated SCFT. Thus, it appears the $F_4$ rank-one instanton VOA should not correspond to a 4d theory. The solicitous reader may question whether such gaugings would actually be possible from the 4d perspective, as there could be a mismatch in the global anomalies for these would-be 4d $Sp(n)$ gaugings. However, we could always glue to hybrid cylinders in the $A_{2n}$ theory as well, which would have a global anomaly\cite{Tachikawa:2018rgw}, and these BRST gaugings would still result in the VOAs of isolated SCFTs. We will see some specific examples of this in the next subsection.
\begin{figure}
\centering
\begin{tikzpicture}\draw[radius=40pt,fill=lightmauve] (-4,0) circle;
\draw[radius=2pt,fill=white]  (-4.5,.9) circle node[below=2pt] {$[2,1^2]$};
\draw[radius=2pt,fill=white]  (-4.5,-1) circle node[above=2pt] {$[2,1^2]$};
\draw[radius=2pt,fill=white]  (-3,0) circle node[left=2pt] {$[1^{4}]$};
\draw[radius=40pt,fill=lightred] (.6,0) circle;
\draw[radius=2pt,fill=white]  (1,.9) circle node[left=2pt] {$\widebar{[1^6]}$};
\draw[radius=2pt,fill=white]  (1,-1) circle node[left=2pt] {$[4]$};
\draw[radius=2pt,fill=white]  (-.5,0) circle node[right=2pt] {$[1^{4}]$};
\path
(-2.9,0)  edge node[above] {$Sp(2)$} (-.6,0);
\node at (-4,-2) {$[(F_4)_5+\frac{1}{2}(4)]$};
\node at (4.4,-2) {$[Spin(7)_8\times SU(2)_5^2]$};
\node at (.5,-2) {$\frac{1}{2}(4,6)$};
\node at (2.4,0) {$\cong$};
\draw[radius=40pt,fill=lightblue] (4.4,0) circle;
\draw[radius=2pt,fill=white]  (3.9,.9) circle node[below=2pt] {$[2,1^2]$};
\draw[radius=2pt,fill=white]  (3.9,-1) circle node[above=2pt] {$[2,1^2]$};
\draw[radius=2pt,fill=white]  (5.4,0) circle node[left=2pt] {$\widebar{[1^{6}]}$};
\end{tikzpicture}
\caption{Duality resulting in the $Spin(7)_8 \times SU(2)_5^2$ VOA. A bar over a nilpotent orbit indicates an untwisted puncture.}
\end{figure}
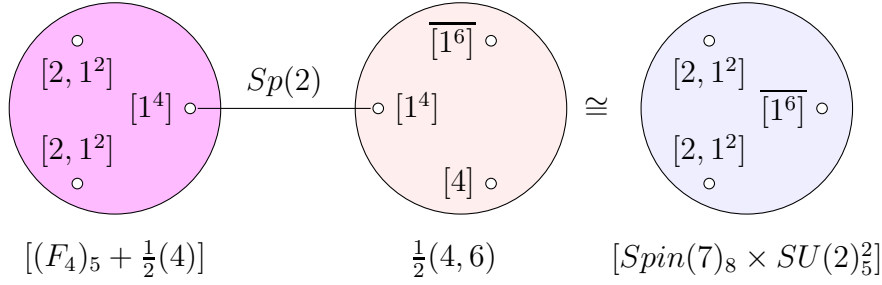

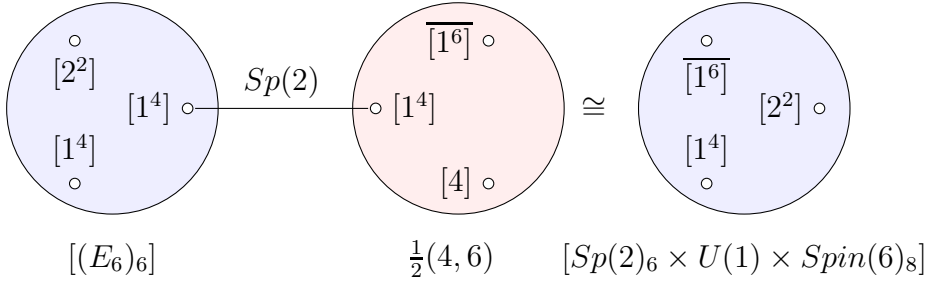
\begin{figure}
\centering
\begin{tikzpicture}\draw[radius=40pt,fill=lightblue] (-4,0) circle;
\draw[radius=2pt,fill=white]  (-4.5,.9) circle node[below=2pt] {$[2^2]$};
\draw[radius=2pt,fill=white]  (-4.5,-1) circle node[above=2pt] {$[1^4]$};
\draw[radius=2pt,fill=white]  (-3,0) circle node[left=2pt] {$[1^4]$};
\draw[radius=40pt,fill=lightred] (.6,0) circle;
\draw[radius=2pt,fill=white]  (1,.9) circle node[left=2pt] {$\widebar{[1^6]}$};
\draw[radius=2pt,fill=white]  (1,-1) circle node[left=2pt] {$[4]$};
\draw[radius=2pt,fill=white]  (-.5,0) circle node[right=2pt] {$[1^{4}]$};
\path
(-2.9,0)  edge node[above] {$Sp(2)$} (-.6,0);
\node at (-4,-2) {$[(E_6)_6]$};
\node at (4.4,-2) {$[Sp(2)_6 \times U(1) \times Spin(6)_8]$};
\node at (.5,-2) {$\frac{1}{2}(4,6)$};
\node at (2.4,0) {$\cong$};
\draw[radius=40pt,fill=lightblue] (4.4,0) circle;
\draw[radius=2pt,fill=white]  (3.9,.9) circle node[below=2pt] {$\widebar{[1^6]}$};
\draw[radius=2pt,fill=white]  (3.9,-1) circle node[above=2pt] {$[1^4]$};
\draw[radius=2pt,fill=white]  (5.4,0) circle node[left=2pt] {$[2^2]$};
\end{tikzpicture}
\caption{Duality resulting in the $Sp(2)_6 \times U(1) \times Spin(6)_8$ VOA. A bar over a nilpotent orbit indicates an untwisted puncture.}
\label{Unresolved}
\end{figure}

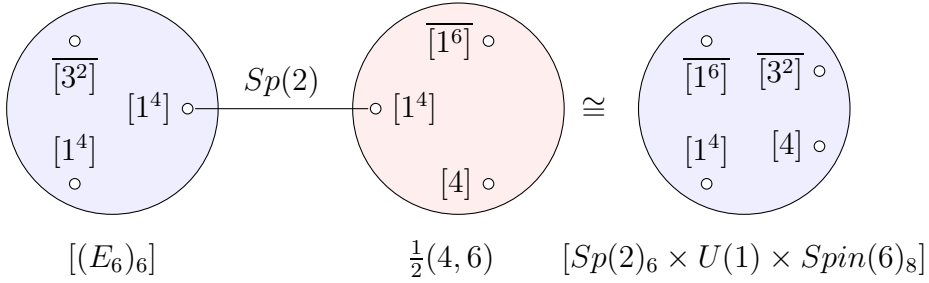
\begin{figure}
\centering
\begin{tikzpicture}\draw[radius=40pt,fill=lightblue] (-4,0) circle;
\draw[radius=2pt,fill=white]  (-4.5,.9) circle node[below=2pt] {$\widebar{[3^2]}$};
\draw[radius=2pt,fill=white]  (-4.5,-1) circle node[above=2pt] {$[1^4]$};
\draw[radius=2pt,fill=white]  (-3,0) circle node[left=2pt] {$[1^4]$};
\draw[radius=40pt,fill=lightred] (.6,0) circle;
\draw[radius=2pt,fill=white]  (1,.9) circle node[left=2pt] {$\widebar{[1^{6}]}$};
\draw[radius=2pt,fill=white]  (1,-1) circle node[left=2pt] {$[4]$};
\draw[radius=2pt,fill=white]  (-.5,0) circle node[right=2pt] {$[1^4]$};
\path
(-2.9,0)  edge node[above] {$Sp(2)$} (-.6,0);
\node at (-4,-2) {$[(E_6)_6]$};
\node at (4.4,-2) {$[Sp(2)_6 \times U(1) \times Spin(6)_8]$};
\node at (.5,-2) {$\frac{1}{2}(4,6)$};
\node at (2.4,0) {$\cong$};
\draw[radius=40pt,fill=lightblue] (4.4,0) circle;
\draw[radius=2pt,fill=white]  (3.9,.9) circle node[below=2pt] {$\widebar{[1^6]}$};
\draw[radius=2pt,fill=white]  (3.9,-1) circle node[above=2pt] {$[1^4]$};
\draw[radius=2pt,fill=white]  (5.4,.5) circle node[left=2pt] {$\widebar{[3^2]}$};
\draw[radius=2pt,fill=white]  (5.4,-.5) circle node[left=2pt] {$[4]$};
\end{tikzpicture}
\caption{Alternative duality frame for the $Sp(2)_6 \times U(1) \times Spin(6)_8$ VOA seen after resolving the $[2^2]$ puncture into $\widebar{[3^2]}$ and $[4]$. A bar over a nilpotent orbit indicates an untwisted puncture.}
\label{Resolved}
\end{figure}

Now the perceptive reader might ask, how do we find VOAs corresponding to known 4d SCFTs given that a BRST gauging of them would result in a three punctured sphere VOA of a twisted class-S theory? The answer to this lies in the discussion earlier on how the VOAs corresponding to known 4d SCFTs have a puncture that in the twisted theory is an atypical puncture, which is actually two punctures in disguise. Thus, performing the above exercise gives the VOA of a three-punctured sphere class-S theory with an atypical puncture, indicating the theory has a non-trivial conformal manifold, which is perfectly consistent. As an example, suppose we take the $(E_6)_6$ VOA found in the $C_2$ theory and glue it to the hybrid cylinder VOA. We then get the class-S VOA corresponding to the sphere with punctures $[1^4],[2^2]$ and $\widebar{[1^6]}$. We can then resolve the $[2^2]$ puncture into the punctures $[4]$ and $\widebar{[3^2]}$. This same VOA duality can then be seen with twisted class-S VOAs, as shown in Figures \ref{Unresolved} and \ref{Resolved}. Going to another duality frame, one sees this is in fact a $Spin(6)$ gauge theory with four hypers in the fundamental and two in the vector. The Hitchin system of the unresolved three punctured sphere is the Hitchin system of the four punctured sphere at a particular point in the conformal manifold. However, the VOA does not depend on the conformal manifold, hence both VOAs are isomorphic.

\subsubsection{Twisted Trinions}\label{TwistedTrinion}

As mentioned earlier, we can use these hybrid gluings to construct the twisted class-S trinions. We have seen the characters are the same and it's trivial to see the current algebra levels are equal. For the central charges to be the same we need
\begin{equation*}
    c_{\text{trin}}+c_{bc}+c_{\text{hybrid}}=c_{\text{twisted}} \to c_{\text{trin}}+c_{bc}=c_{\text{twisted}}-c_{\text{hybrid}}.
\end{equation*} The R.H.S. of the second equation is just the change in the central charge from the principal Drinfeld-Sokolov reduction of a $\mathfrak{g}$ current algebra. From \cite{Arakawa:2018egx} the combined central charge of the non-simply laced trinion and the $bc$ ghost systems is \begin{equation*}
    \big(3\dim \mathfrak{g} -\text{rank } \mathfrak{g}-24(\rho| \rho^{\vee}) \big)-\big(2 \dim \mathfrak{g}) = \dim \mathfrak{g} -\text{rank } \mathfrak{g}-24(\rho| \rho^{\vee})
\end{equation*}
where $\rho$ and $\rho^{\vee}$ are half the sum of roots and coroots respectively. The R.H.S. is precisely equal to the change in central charge from the principal Drinfeld-Sokolov reduction of a $\mathfrak{g}$ current algebra at the critical level, as expected. In fact, using the technology of Feigin-Frenkel gluing, we can directly see they are the same, at least in the $\mathbb{Z}_2$ and $\mathbb{Z}_3$ twisted case. We will now provide an extremely brief review of the construction of class-S VOAs in \cite{Arakawa:2018egx}.

Consider the equivariant affine W-algebra $\mathbf{W}_G$. From the class-S TQFT perspective this is the VOA associated to a cap and notably has a $\mathfrak{g}$ current algebra at the critical level. Recall that a current algebra at the critical level has a center deemed the Feigin-Frenkel center. We can tensor $n$ of these $\mathbf{W}_G$ together and identify their Feigin-Frenkel centers via a BRST gauging. This then gives the genus zero class-S VOA with $n$ full punctures which is denoted $V_{G,n}$. Thus, these class-S VOAs all have a single Feigin-Frenkel center, and may be considered modules over it. We will denote this Feigin-Frenkel gluing with $*$. One can also consider taking a product of two class-S VOAs, and then gauging the diagonal current algebra. We denote this operation with a $\circ$. It was additionally shown by Arakawa that \begin{equation}
    V_{G,n} \circ V_{G,m} = V_{G,m+n-2}.
\end{equation}

When the corresponding Lie algebra has an outer automorphism, \cite{Beem:2022mde} showed there is a mixed Feigin-Frenkel gluing between modules for the $J$ and $G$ type Feigin-Frenkel centers. In particular the twisted trinion was found to be $W_J *_u (W_G *_t W_G)$  where $*_u$ denotes the mixed gluing, and $*_t$ denotes the ordinary Feigin-Frenkel gluing of type $\mathfrak{g}$. 

The hybrid cylinder VOA was noted in \cite{Beem:2022mde} to be the mixed gluing of the type $\mathbf{W}_G$ and $\mathbf{W}_J$ cap algebras. We can also make use of their Lemma 4.12, which under some hypotheses\footnote{Specifically, we need $M$ to be free over a subalgebra of the Feigin-Frenkel center called $\mathcal{Z}_{u,<0}$ and for $U$ to be semijective in a certain subcategory of modules over the corresponding Feigin-Frenkel center. For $M=\mathbf{W}_J$ and $U=V_{G,n}$, these respective properties were shown in \cite{Arakawa:2018egx}} on $M,U,$ and $V$ gives \begin{equation*}
    M *_u (U \circ_t V) = (M *_u U) \circ_t V
\end{equation*} where $\circ_t$ indicates gauging the corresponding diagonal $\mathfrak{g}$ current algebra at 2d level $-2h^{\vee}$. 
Thus, when we glue the hybrid cylinder to a non-simply laced trinion we have
\begin{equation}
    (V_{J,1} *_u V_{G,1}) \circ_t V_{G,3} = V_{J,1} *_u (V_{G,1} \circ_t V_{G,3}) = V_{J,1} *_u V_{G,2}.
\end{equation}
The R.H.S is just the definition of the twisted class-S trinion in \cite{Beem:2022mde}. 

For certain class-S types this procedure does not seem very useful, though for others the VOA glued to the non-simply laced trinion is quite simple. For the $D_n$ case, we can always construct the twisted trinion by adding some hypermultiplets and gluing to the $C_{n-1}$ trinion. From \cite{Chacaltana:2014ica} we know the hybrid cylinder is a half-hyper in the bifundamental of $Sp(n-1)$ and $Spin(2n)$. In total the number of free hypers is $2n^2-2n$. At the level of symplectic varieties we then have that the twisted trinion Higgs branch can be obtained from the hyperkh\"{a}ler quotient of the product of the corresponding Moore-Tachikawa variety and $\mathbb{H}^{2n^2-2n}$. 

In the $D_4$ case the automorphism group is larger so there are three trinions to construct. The first is the trinion of the $(1,\omega, \omega^2)$ twisted sector. The trinion has two punctures labeled by $G_2$ nilpotent orbits and one labeled by a $D_4$ nilpotent orbit. From \cite{Chacaltana:2016shw} we know the VOA we would like to glue to the $G_2$ trinion is the $Sp(3)_{16}$ gauging of the product theory $(E_7)_8+ \frac{1}{2}(4,8)$. This hybrid cylinder is just the mixed gluing of the $G_2$ and $D_4$ equivariant affine W-algebras.

There are an additional two trinions that were not constructed in \cite{Beem:2022mde}, one of which is the trinion VOA of the $(\omega,\omega,\omega)$ twisted sector in the $\mathbb{Z}_3$ twisted $D_4$ theory. Consider the fixture with two full punctures and a regular orbit puncture in the $(\omega,\omega,\omega)$ sector. Resolving the $G_2$ puncture for this reduced VOA shows this fixture is given by the gauging of the diagonal $Sp(3)_{16}$ of the product of two $(E_7)_8$ theories, hence the VOA can then be obtained via the corresponding BRST reduction. We propose that the twisted trinion is obtained by gluing this VOA to the $G_2$ trinion, as shown in Figure \ref{G2}.

\begin{figure}
\centering
\begin{tikzpicture}\draw[radius=40pt,fill=lightblue] (-4,0) circle;
\draw[radius=2pt,fill=white]  (-4.5,.9) circle node[below=2pt] {$0$};
\draw[radius=2pt,fill=white]  (-4.5,-1) circle node[above=2pt] {$0$};
\draw[radius=2pt,fill=white]  (-3,0) circle node[left=2pt] {$0$};
\draw[radius=40pt,fill=lightblue] (.6,0) circle;
\draw[radius=2pt,fill=white]  (1,.9) circle node[left=2pt] {$0$};
\draw[radius=2pt,fill=white]  (1,-1) circle node[left=2pt] {$G_2$};
\draw[radius=2pt,fill=white]  (-.5,0) circle node[right=2pt] {$0$};
\path
(-2.9,0)  edge node[above] {$G_2$} (-.6,0);
\node at (-4,-2) {$[(G_2)_8^3]$};
\node at (4.4,-2) {$[(G_2)_8^3]$};
\node at (.5,-2) {$[(G_2)_8^2]$};
\node at (2.4,0) {$\cong$};
\draw[radius=40pt,fill=lightblue] (4.4,0) circle;
\draw[radius=2pt,fill=white]  (3.9,.9) circle node[below=2pt] {$0$};
\draw[radius=2pt,fill=white]  (3.9,-1) circle node[above=2pt] {$0$};
\draw[radius=2pt,fill=white]  (5.4,0) circle node[left=2pt] {$0$};
\end{tikzpicture}
\caption{Duality resulting in the trinion for the $(\omega, \omega, \omega)$ twisted sector. The VOA on the left side of the gauging is the $G_2$ trinion and the fixture on the right is $Sp(3)_{16}$ gauging of two $(E_7)_8$ VOAs. }
\label{G2trin1}
\end{figure}
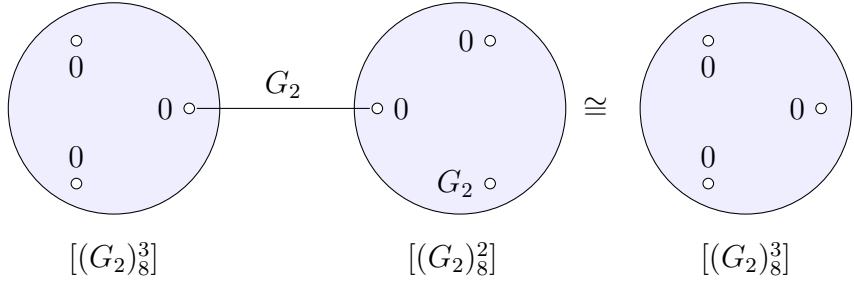

\begin{figure}
\centering
\begin{tikzpicture}\draw[radius=40pt,fill=lightblue] (-4,0) circle;
\draw[radius=2pt,fill=white]  (-4.5,.9) circle node[below=2pt] {$[1^6]$};
\draw[radius=2pt,fill=white]  (-4.5,-1) circle node[above=2pt] {$[1^6]$};
\draw[radius=2pt,fill=white]  (-3,0) circle node[left=2pt] {$[1^6]$};
\draw[radius=40pt,fill=lightblue] (.6,0) circle;
\draw[radius=2pt,fill=white]  (1,.9) circle node[left=2pt] {$[6]$};
\draw[radius=2pt,fill=white]  (1,-1) circle node[left=2pt] {$0$};
\draw[radius=2pt,fill=white]  (-.5,0) circle node[right=2pt] {$[1^6]$};
\path
(-2.9,0)  edge node[above] {$Sp(3)$} (-.6,0);
\node at (-4,-2) {$[Sp(3)_8^3]$};
\node at (4.4,-2) {$[Sp(3)_8^2 \times (G_2)_8]$};
\node at (.5,-2) {$[(E_7)_8]$};
\node at (2.4,0) {$\cong$};
\draw[radius=40pt,fill=lightblue] (4.4,0) circle;
\draw[radius=2pt,fill=white]  (3.9,.9) circle node[below=2pt] {$[1^6]$};
\draw[radius=2pt,fill=white]  (3.9,-1) circle node[above=2pt] {$[1^6]$};
\draw[radius=2pt,fill=white]  (5.4,0) circle node[left=2pt] {$0$};
\end{tikzpicture}
\caption{Duality resulting in the trinion for the non-abelian twisted sector VOA.}
\label{G2trin2}
\end{figure}
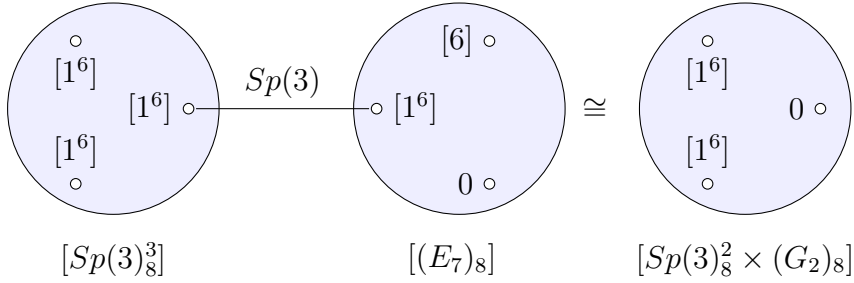

We can use a similar method to construct the VOAs for the non-abelian twisted $D_4$ theory. Suppose we take the $C_3$ trinion and glue it to the $(E_7)_8$ theory. This corresponds to the non-abelian twisted fixture with punctures $[1^6],[6]$ and $0$ and thus the gluing should give a VOA whose character agrees with the trinion in the non-commuting twist sector. We can also check that the 2d central charge of the VOA is -298, as we would expect from the 4d invariants. Interestingly, we already have a realization of the $(E_7)_8$ theory amongst the $C_3$ fixtures, and we see that the non-abelian twisted trinion corresponds to the $C_3$ four punctured sphere with punctures $[1^6],[1^6],[2^3],[3^2]$.

This construction appears to explain the strange phenomenon in the non-abelian twisted sector where two different weakly coupled limits appear at the same point of the conformal manifold, at least at the VOA level. Suppose we have a twisted VOA with four $C_3$ punctures. This can be obtained by gluing together either two fixtures in the non-abelian twisted sector or two fixtures in the $\mathbb{Z}_2$ twisted sector. This follows from the fact that we are connecting two $C_3$ trinions via a VOA with a $Sp(3)_8^2$ symmetry that arises in two different ways. If we can show that the two constructions of this $Sp(3)_8^2$ VOA are equal then this proves the additional duality frames. The first construction of this $Sp(3)_8^2$ VOA involves gauging the diagonal $(G_2)_{16}$ of the product of two $(E_7)_8$ theories. The second involves gauging the $Spin(8)_{24}$ of 48 spin $\frac{1}{2}$ $\beta\gamma$ systems. It's easy to see the central charges are the same and from the identifications of \cite{Distler:2021cwz} we expect the characters to be the same, though an actual proof of the equivalence of these VOAs escapes the author. 

The $E_7$ current algebra at level $-4$ seems to play a fundamental role in the $D_4$ case. This appears to be due to the fact that it contains a commuting pair of $Sp(3)$ and $G_2$ current algebras at their critical levels. It's tempting to conjecture that it arises as a mixed gluing of the $G_2$ and $C_3$ equivariant affine W-algebras. We mention that in principle a similar construction could also be done for the twisted $A_{n}$ and $E_6$ VOAs. However, the VOAs typically do not have alternative constructions. An exception to this statement is the hybrid cylinder for the $A_2$ theory, which is the rank-two $SU(3)$ instanton theory. 

\subsection{Gluing Irregular Punctures}
The above story presumably generalizes to gaugings that are not necessarily the entire flavour symmetry of a full puncture. That is, we take a class-S VOA tensored with another VOA, and gauge some manifest symmetry that is not the full symmetry from the full puncture. In the class-S literature this is known to be equivalent to taking a different class-S theory with two punctures colliding, see \cite{Gaiotto:2011xs}. When the punctures collide, they bubble off to form a new theory called an irregular fixture, which is coupled to the rest of the class-S theory via some exactly marginal gauging. If $\rho_1$ and $\rho_2$ denote the $SU(2)$ embeddings of the two punctures that collide to produce the puncture with embedding $\rho_3$, then coupling the irregular fixture via gauging changes the index in the following manner: 
\begin{equation}
   \sum_{\mathfrak{R}} ...\psi^{\rho_3}_{\mathfrak{R}}(\tau) \to \sum_{\mathfrak{R}}... \frac{\psi^{\rho_1}_{\mathfrak{R}}(\tau)\psi^{\rho_2}_{\mathfrak{R}}(\tau)}{\psi^{\rho}_{\mathfrak{R}}(\tau)}.
\end{equation}

Since the preceding factors in each term in the sum do not affect the change in the character, we expect something similar to work with these non-simply laced VOAs. It would be nice to have a rigorous understanding of these irregular gluings, but we leave such an investigation for future work. Note when gluing to a twisted puncture, we get the OPE of a twisted and an untwisted puncture. When the resulting twisted puncture corresponds to the regular orbit, we then expect to get the VOA of a twisted class-S fixture when gluing the irregular fixture to a non-simply laced fixture.

For example, in the $C_2$ theory, adding three hypermultiplets in the fundamental of the $SU(2)$ for the $[2,1^2]$ punctures gives the OPE of the twisted puncture $[4]$ and the untwisted puncture $[2^2]$. Let us perform this same operation with the $C_2$ theory, in particular the $F_{4}$ instanton fixture. We should get a VOA with symmetry $Sp(3)_5\times SU(2)_8$ and central charge $c_{2d}=-29$ plus two $\beta\gamma$ systems. At the level of indices we have
\begin{equation}
 \sum_{\mathfrak{R}}\frac{\psi^{[1^4]}_{\mathfrak{R}}\psi^{[2,1^2]}_{\mathfrak{R}}\psi^{[2,1^2]}_{\mathfrak{R}}}{\psi^{[4]}_{\mathfrak{R}}} \to \sum_{\mathfrak{R}}\frac{\psi^{[1^4]}_{\mathfrak{R}}\psi^{[2,1^2]}_{\mathfrak{R}}}{\psi^{[4]}_{\mathfrak{R}}}\times \frac{\psi^{[4]}_{\mathfrak{R}}\tilde{\psi}^{[3,1^3]}_{\mathfrak{R}}}{\tilde{\psi}^{[5,1]}_{\mathfrak{R}}}=\sum_{\mathfrak{R}} \frac{\psi^{[1^4]}_{\mathfrak{R}}\psi^{[2,1^2]}_{\mathfrak{R}}\tilde{\psi}^{[3,1^3]}_{\mathfrak{R}}}{\tilde{\psi}^{[5,1]}_{\mathfrak{R}}}.
\end{equation}
Thus, we just expect the VOA of a twisted class-S fixture. This fixture was considered in \cite{Chacaltana:2012ch} and was found to be the rank-one $Sp(3)_5\times SU(2)_8$ SCFT in addition to two free hypers, in exact agreement with our prediction from the gauging of the $F_4$ instanton VOA. At the level of hyperkh{\"a}ler quotients, the identification of the $SU(2)$ quotient of the minimal nilpotent orbit of $F_4$ times $\mathbb{H}^3$ as the Higgs branch of the $Sp(3)_5\times SU(2)_8$ SCFT was conjectured in \cite{Bourget:2020asf}. We could perform a similar operation on the $Sp(n)_{n+2}\times SU(2)_5$ VOAs we found to obtain the $Sp(n)_{n+2}\times SU(2)_8$ VOAs found in \cite{Chacaltana:2013oka}.
\begin{figure}
\centering
\begin{tikzpicture}\draw[radius=40pt,fill=lightmauve] (-4,0) circle;
\draw[radius=2pt,fill=white]  (-4.5,.9) circle node[below=2pt] {$[2,1^2]$};
\draw[radius=2pt,fill=white]  (-4.5,-1) circle node[above=2pt] {$[1^4]$};
\draw[radius=2pt,fill=white]  (-3,0) circle node[left=2pt] {$[2,1^2]$};
\draw[radius=40pt,fill=lightred] (.6,0) circle;
\draw[radius=2pt,fill=white]  (1,.9) circle node[left=2pt] {$\widebar{[3,1^3]}$};
\draw[radius=2pt,fill=white]  (1,-1) circle node[left=2pt] {$[4]$};
\draw[radius=2pt,fill=white]  (-.5,0) circle node[right=2pt] {$([2,1^{2}],SU_2)$};
\path
(-2.9,0)  edge node[above] {$SU(2)$} (-.6,0);
\node at (-4,-2) {$[(F_4)_5+\frac{1}{2}(4)]$};
\node at (4.4,-2) {$[Sp(3)_5\times SU(2)_8+\frac{1}{2}(4)]$};
\node at (.5,-2) {$\frac{1}{2}(2,3)$};
\node at (2.4,0) {$\cong$};
\draw[radius=40pt,fill=lightmauve] (4.4,0) circle;
\draw[radius=2pt,fill=white]  (3.9,.9) circle node[below=2pt] {$[2,1^2]$};
\draw[radius=2pt,fill=white]  (3.9,-1) circle node[above=2pt] {$[1^4]$};
\draw[radius=2pt,fill=white]  (5.4,0) circle node[left=2pt] {$\widebar{[3,1^3]}$};
\end{tikzpicture}
\caption{Duality resulting in the $Sp(3)_5 \times SU(2)_8$ VOA. A bar over a nilpotent orbit indicates an untwisted puncture.}
\end{figure}
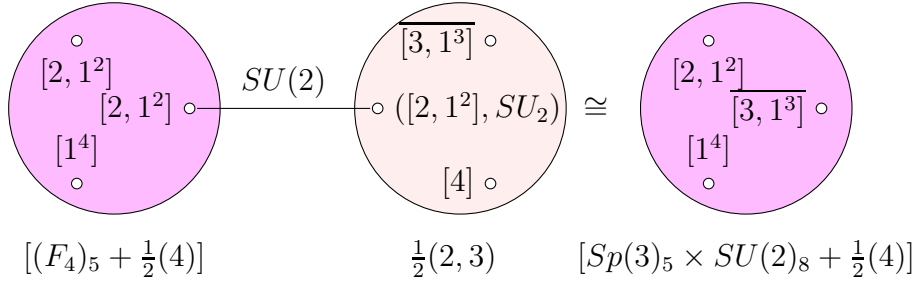

At the VOA level we could also consider BRST gaugings that glue various twisted $A_{2n}$ VOAs to these non-simply laced VOAs. For example, we could consider gauging the diagonal $SU(2)_8$ of the tensor product of the $F_4$ and $SU(3)$ rank-one instanton VOAs. This results in a VOA with $Sp(3)_5\times U(1)$ flavour symmetry and central charge $c_{2d}=-34$. A rank-two theory with these invariants has been studied in \cite{Giacomelli:2020gee, Martone:2021drm ,Martone:2021ixp,Zafrir:2016wkk}. Using the $C_2$ realization of the $F_4$ intstanton VOA and the known irregular fixture that corresponds to gluing the $SU(3)$ instanton theory to the $[2,1^2]$ puncture in the twisted $A_4$ theory, we obtain the rank-two theory with $Sp(3)_5\times U(1)$ flavour symmetry in the twisted $A_4$ theory\cite{DistlerElliotWIP2}.
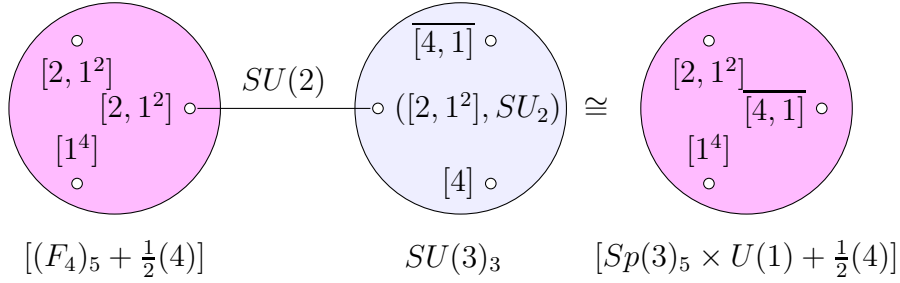
\begin{figure}
\centering
\begin{tikzpicture}\draw[radius=40pt,fill=lightmauve] (-4,0) circle;
\draw[radius=2pt,fill=white]  (-4.5,.9) circle node[below=2pt] {$[2,1^2]$};
\draw[radius=2pt,fill=white]  (-4.5,-1) circle node[above=2pt] {$[1^4]$};
\draw[radius=2pt,fill=white]  (-3,0) circle node[left=2pt] {$[2,1^2]$};
\draw[radius=40pt,fill=lightblue] (.6,0) circle;
\draw[radius=2pt,fill=white]  (1,.9) circle node[left=2pt] {$\widebar{[4,1]}$};
\draw[radius=2pt,fill=white]  (1,-1) circle node[left=2pt] {$[4]$};
\draw[radius=2pt,fill=white]  (-.5,0) circle node[right=2pt] {$([2,1^{2}],SU_2)$};
\path
(-2.9,0)  edge node[above] {$SU(2)$} (-.6,0);
\node at (-4,-2) {$[(F_4)_5+\frac{1}{2}(4)]$};
\node at (4.4,-2) {$[Sp(3)_5\times U(1)+\frac{1}{2}(4)]$};
\node at (.5,-2) {$SU(3)_3$};
\node at (2.4,0) {$\cong$};
\draw[radius=40pt,fill=lightmauve] (4.4,0) circle;
\draw[radius=2pt,fill=white]  (3.9,.9) circle node[below=2pt] {$[2,1^2]$};
\draw[radius=2pt,fill=white]  (3.9,-1) circle node[above=2pt] {$[1^4]$};
\draw[radius=2pt,fill=white]  (5.4,0) circle node[left=2pt] {$\widebar{[4,1]}$};
\end{tikzpicture}
\caption{Duality resulting in the $Sp(3)_5 \times U(1)$ VOA. A bar over a nilpotent orbit indicates an untwisted puncture.}
\end{figure}
There exists a similar construction to get the rest of the $Sp(n)_{n+2}\times U(1)$ series found in \cite{Chacaltana:2014nya,Zafrir:2016wkk} by gluing to the $Sp(n)_{n+2}\times SU(2)_5$ series. As another example, one could take the twisted $A_2$ realization of the $SU(3)$ instanton theory\cite{Beem:2020pry} and glue one of the twisted full punctures to the $A_1$ trinion. This resulting VOA should have $Sp(2)_4 \times U(1)$ flavour symmetry and central charge $c_{2d}=-19$. This agrees precisely with the Argyres-Wittig theory \cite{Argyres:2007tq}. 

In these two examples, from the 4d perspective these gaugings are inconsistent due to the presence of Witten's global anomaly in the twisted $A_{2n}$ theory \cite{Tachikawa:2018rgw}. We are also unsure if there is a 3d interpretation, though we mention the mismatch of the global anomalies in 4d would result in the corresponding 3d Yang-Mills field having a non-zero Chern-Simons level \cite{Alvarez-Gaume:1983ihn}.

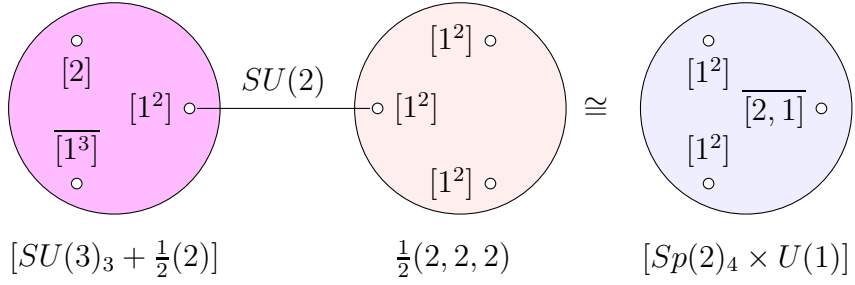
\begin{figure}
\centering
\begin{tikzpicture}\draw[radius=40pt,fill=lightmauve] (-4,0) circle;
\draw[radius=2pt,fill=white]  (-4.5,.9) circle node[below=2pt] {$[2]$};
\draw[radius=2pt,fill=white]  (-4.5,-1) circle node[above=2pt] {$\widebar{[1^3]}$};
\draw[radius=2pt,fill=white]  (-3,0) circle node[left=2pt] {$[1^2]$};
\draw[radius=40pt,fill=lightred] (.6,0) circle;
\draw[radius=2pt,fill=white]  (1,.9) circle node[left=2pt] {$[1^2]$};
\draw[radius=2pt,fill=white]  (1,-1) circle node[left=2pt] {$[1^2]$};
\draw[radius=2pt,fill=white]  (-.5,0) circle node[right=2pt] {$[1^{2}]$};
\path
(-2.9,0)  edge node[above] {$SU(2)$} (-.6,0);
\node at (-4,-2) {$[SU(3)_3+\frac{1}{2}(2)]$};
\node at (4.4,-2) {$[Sp(2)_4\times U(1)]$};
\node at (.5,-2) {$\frac{1}{2}(2,2,2)$};
\node at (2.4,0) {$\cong$};
\draw[radius=40pt,fill=lightblue] (4.4,0) circle;
\draw[radius=2pt,fill=white]  (3.9,.9) circle node[below=2pt] {$[1^2]$};
\draw[radius=2pt,fill=white]  (3.9,-1) circle node[above=2pt] {$[1^2]$};
\draw[radius=2pt,fill=white]  (5.4,0) circle node[left=2pt] {$\widebar{[2,1]}$};
\end{tikzpicture}
\caption{Duality resulting in the $Sp(2)_4 \times U(1)$ VOA. A bar over a nilpotent orbit indicates an untwisted puncture.}
\end{figure}

Let's look at another example, this time in the $B_3$ theory. Take our realization in the $B_3$ VOAs of the rank-two $F_4$ instanton VOA. It has an $SU(2)_6$ flavour symmetry which we could consider gauging by adding two half-hypermultiplets in the fundamental. This would then give us an $(F_4)_{10} \times U(1)$ VOA with central charge $c_{2d}=-64$. This gauging in the twisted $A_5$ theory replaces the $[3,2^2]$ puncture with a $[7]$ and untwisted $[4,2]$ puncture. The resulting theory is a rank-two 4d SCFT with flavour symmetry $(F_4)_{10}\times U(1)$ and Coulomb branch generators of dimensions four and five which matches with the properties of the rank-two theory found in \cite{Distler:2017xba,Wang:2018gvb,Giacomelli:2020gee}.
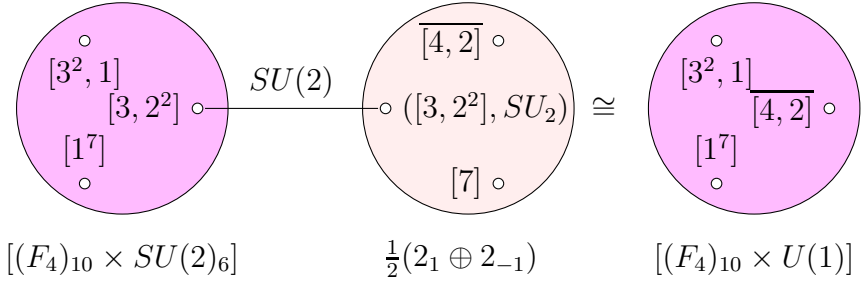
\begin{figure}
\centering
\begin{tikzpicture}\draw[radius=40pt,fill=lightmauve] (-4,0) circle;
\draw[radius=2pt,fill=white]  (-4.5,.9) circle node[below=2pt] {$[3^2,1]$};
\draw[radius=2pt,fill=white]  (-4.5,-1) circle node[above=2pt] {$[1^7]$};
\draw[radius=2pt,fill=white]  (-3,0) circle node[left=2pt] {$[3,2^2]$};
\draw[radius=40pt,fill=lightred] (.6,0) circle;
\draw[radius=2pt,fill=white]  (1,.9) circle node[left=2pt] {$\widebar{[4,2]}$};
\draw[radius=2pt,fill=white]  (1,-1) circle node[left=2pt] {$[7]$};
\draw[radius=2pt,fill=white]  (-.5,0) circle node[right=2pt] {$([3,2^2],SU_2)$};
\path
(-2.9,0)  edge node[above] {$SU(2)$} (-.6,0);
\node at (-4,-2) {$[(F_4)_{10} \times SU(2)_6]$};
\node at (4.4,-2) {$[(F_4)_{10} \times U(1)]$};
\node at (.5,-2) {$\frac{1}{2}(2_1\oplus 2_{-1})$};
\node at (2.4,0) {$\cong$};
\draw[radius=40pt,fill=lightmauve] (4.4,0) circle;
\draw[radius=2pt,fill=white]  (3.9,.9) circle node[below=2pt] {$[3^2,1]$};
\draw[radius=2pt,fill=white]  (3.9,-1) circle node[above=2pt] {$[1^7]$};
\draw[radius=2pt,fill=white]  (5.4,0) circle node[left=2pt] {$\widebar{[4,2]}$};
\end{tikzpicture}
\caption{Duality resulting in the $(F_4)_{10} \times U(1)$ VOA. A bar over a nilpotent orbit indicates an untwisted puncture.}
\end{figure}

We also mention that this same argument can be made for the $Sp(2)_{14}\times SU(2)_5$ VOA found amongst the $G_2$ VOAs. We could consider gluing three free hypers and gauging the diagonal $SU(2)_8$ to get the rank-two $Sp(2)_{14}\times SU(2)_8$ theory which is an isolated SCFT.

\subsection{Gauging Outside Class-S}\label{Outside}
Somewhat suggestively, there also exists a rank-two theory with flavour symmetry $(G_2)_{\frac{20}{3}}$, which has the same level as the $G_2$ symmetry of the rank-two $G_2$ instanton VOA. It's tempting to conjecture that its VOA arises in a similar manner to the $(F_4)_{10}\times U(1)$ theory. In fact, there is an obvious candidate. Suppose we take the product of the rank-two $G_2$ instanton VOA with the rank-one $SU(2)$ instanton VOA and a free spin 1/2 $\beta \gamma$ system. Recall that the 4d symmetry of a the two $G_2$ instanton theory also has an $SU(2)_{\frac{13}{3}}$ factor. We can gauge a diagonal $SU(2)$ at level $\frac{13}{3}+\frac{8}{3}+\frac{3}{3}=8$ to obtain a theory with $(G_2)_{\frac{20}{3}}$ symmetry. We can additionally calculate the central charge of such a theory and find $c_{2d}=-44$. This agrees precisely with the theory $\widehat{\mathcal{T}}_{D_4,3}$ found in \cite{Giacomelli:2020gee}. Furthermore, computing the dimension of the corresponding hyperkh{\"a}ler quotient, we find the answer is six, which agrees with the dimension of the Higgs branch of the four-dimensional theory. Computing the Schur index to order $\tau^7$ gives 
\begin{equation*}
    1+14\tau^2+120\tau^4+736\tau^6+O(\tau^8)
\end{equation*}
though we know of no existing calculations in the literature to compare to. 

We could modify the above set-up by performing a highest weight nilpotent Higgsing/Drinfeld-Sokolov reduction of the $(G_2)_{\frac{20}{3}}$ symmetry. Note that the BRST differential for the Drinfeld-Sokolov reduction anticommutes with the BRST differential for gauging the $SU(2)$ current algebra, hence we may perform them in either order. Doing the Drinfeld-Sokolov reduction first, we get the $SU(2)$ gauging of the rank-one $G_2$ instanton theory times the $SU(2)$ rank-one instanton theory with two free hypers. Computing the level of the diagonal $SU(2)$ gives $\frac{10}{3}+\frac{8}{3}+\frac{6}{3}=8$ as expected. The resulting theory should then have a $SU(2)_{10}\times U(1)$ symmetry and 2d central charge $-24$. On the other hand, from \cite{Giacomelli:2020gee} we know that this Higgsing of the $G_2$ symmetry results in the S-fold theory $\mathcal{S}_{A_1,3}^{(1)}$ which is a rank-one theory found in \cite{Argyres:2016xua} and whose data agrees precisely with our prediction. At the level of symplectic varieties, the corresponding hyperkh{\"a}ler quotient was conjectured to be the Higgs branch of the rank one $SU(2)_{10}\times U(1)$ theory in \cite{Bourget:2020asf}. The VOA of this $SU(2)_{10}\times U(1)$ theory was recently constructed in \cite{Beem:2024fom}. Computing the character of our BRST reduction we find 
\begin{equation*}
    1+4\tau^2+8\tau^3+17\tau^4+36\tau^5+77\tau^6+O(\tau^7)
\end{equation*} which agrees with their result. We now find ourselves in a similar situation to what occurred with the rank-one $F_4$ instanton VOA. If the $G_2$ instanton theory existed, we could certainly perform the gauging above, and we would expect to find a theory with a non-trivial conformal manifold. In contrast, we appear to find the VOA of the $SU(2)_{10}\times U(1)$ theory, which is an isolated $\mathcal{N}=2$ SCFT. Thus, it seems the rank-one $G_2$ instanton theory cannot exist. If the reader is concerned about global anomalies, note that if the $G_2$ instanton theory existed, the $SU(2)$ with embedding index one into the $G_2$ cannot have a global anomaly \cite{Shimizu:2017kzs}. Additionally, the $SU(2)$ flavour symmetry of the rank-one $SU(2)$ instanton theory doesn't have a global anomaly and neither does the $SU(2)$ flavour of two half-hypers in the fundamental. Hence, the gauging would be anomaly free.

Let's consider one more example, take the rank-two $SU(2)$ instanton VOA tensored with the $G_2$ rank-one instanton VOA and a free $\beta\gamma$ system/free hyper. This has a 4d symmetry $SU(2)_{\frac{16}{3}}\times SU(2)_{\frac{11}{3}} \times (G_2)_{\frac{10}{3}} \times SU(2)_1$. We could consider gauging the diagonal $SU(2)_8$ and we are left with a VOA with 4d symmetry $SU(2)_{\frac{16}{3}} \times SU(2)_{10}$ and 2d central charge $-38$, which agrees with the data of the rank-two theory $\mathcal{T}_{A_1,3}^{(2)}$ found in \cite{Giacomelli:2020jel}. Additionally the dimension of the hyperkh\"{a}ler quotient is four, which agrees with the dimension of the Higgs branch. Lastly, computing the Hilbert series of the hyperkh\"{a}ler quotient up to order $\tau^{42}$ gives \begin{equation*}\begin{split}
    & 1+6 \tau ^2+4 \tau ^3+24 \tau ^4+34 \tau ^5+80 \tau ^6+134 \tau ^7+246 \tau ^8+406 \tau ^9+682 \tau ^{10}+1046 \tau ^{11}+1656 \tau ^{12}\\ & +2430 \tau ^{13}+3651 \tau ^{14}+5190 \tau ^{15}+7430 \tau ^{16}+10288 \tau ^{17}+14209 \tau ^{18}+19196 \tau ^{19}+25795 \tau ^{20} \\ &+34022 \tau ^{21}+44717 \tau ^{22}+57798 \tau ^{23}+74505 \tau ^{24}+94696 \tau ^{25}+119916 \tau ^{26}+150212 \tau ^{27}+187340 \tau ^{28} \\ &+231618 \tau ^{29}+285114 \tau ^{30}+348254 \tau ^{31}+423798 \tau ^{32}+512108 \tau ^{33}+616803 \tau ^{34}+738244 \tau ^{35}\\ &+880763 \tau ^{36}+1045186 \tau ^{37}+1236352 \tau ^{38}+1455718 \tau ^{39}+1708805 \tau ^{40}+1997422 \tau ^{41}+2328430 \tau ^{42}+O\left(\tau ^{43}\right)
    \end{split}
\end{equation*}
which agrees with the result from the magnetic quiver \cite{Bourget:2020mez}.

We also note that the consistency of the above conjecture under Drinfeld-Sokolov reduction of the $SU(2)_{16/3}$. Performing such a reduction of the VOA should then give the BRST reduction of the product of the rank-one $SU(2)$ and $G_2$ VOAs with two $\beta\gamma$ systems. On the other hand, we know from the stratification of the Higgs branch that this should give the rank-one $SU(2)_{10}\times U(1)$ theory. Thus, we find perfect consistency with our earlier identification.

It's tempting to speculate that there exists a rank-three 4d $\mathcal{N}=2$ SCFT with Couloumb branch parameters of dimension $4,\frac{8}{3}$ and $\frac{10}{3}$ with flavour symmetry $(G_2)_{\frac{20}{3}}\times SU(2)_{\frac{16}{3}}$ whose VOA is obtained via gauging the diagonal $SU(2)_8$ of the product of the rank-two $SU(2)$ and $G_2$ instanton VOAs. Through various choices of Drinfeld-Sokolov reductions we would then obtain the theories and BRST reductions we have seen throughout this subsection. We also note that much like the case of mismatching global anomalies, we are are unsure of a 3d interpretation of these dualities.  

 \section{Discussion and Open Questions}\label{Discuss}

We have observed that non-simply laced class-S VOAs and their associated quiver gauge theories can be quite interesting. Furthermore, various dualities were found among these VOAs that demonstrate that many do not correspond to 4d $\mathcal{N}=2$ theories. More generally, we can eliminate any theory capable of flowing on its Higgs branch to these disallowed VOAs.  It might be worth exploring whether there is any significance in the construction of VOAs for various isolated 4d SCFTs through BRST gaugings of VOAs not associated with 4d theories. As we have seen in Section \ref{Outside}, this phenomenon seems to be a frequent occurrence with 4d $\mathcal{N}=2$ SCFTs, and we hope to more thoroughly explore the construction of VOAs of theories outside of class-S using non-physical BRST gaugings in future work. 

There is another central question: what exact invariant of these 3d SCFTs do these non-simply laced class-S VOAs correspond to, if any? While it was conjectured in \cite{Beem:2022mde} that they are the C-twist boundary VOAs, this does not seem likely in our opinion. If they are not the C-twist boundary VOAs, then what are they? A somewhat unsatisfactory answer to this question involves the free field realizations of \cite{Beem:2019tfp}.  In their work, they found a sort of inverse Drinfeld-Sokolov method to produce the VOA of a 4d theory based on the stratification of its Higgs branch, and applied it in many examples, such as the DC series. It's tempting to conjecture that their construction works for the Coulomb branches of these non-simply laced class-S theories. We have seen a large amount of evidence for this already, as we found $F_4$ instanton VOAs and many VOAs whose Drinfeld-Sokolov reductions result in them. Thus, we could use the stratification of the Higgs/Coulomb branch to define a VOA, but only for a subset of 3d $\mathcal{N}=4$ theories, as there are many with Higgs/Coulomb branches that are not compatible with such a construction. For example, this would not work for theories which have Higgs/Coulomb branches that are minimal nilpotent orbits of Lie algebras outside of $C_n$ and the DC series. Additionally, there are 3d theories with trivial Higgs/Coulomb branch, just as in four dimensions, so this procedure would not work to define a VOA for them. Perhaps this is a hint that there is a more illuminating characterization of these VOAs. 

We have seen how much of the machinery of class-S carries over to the non-simply laced case. The character of the non-simply laced VOAs is given in terms of Arakawa's formula, while the Hilbert series of genus zero theories follows from the 3d perspective. The fact that these both come from twisted wavefunctions seems to suggest an extra grading for the non-simply laced VOAs and hints that the Macdonald index should also apply. It would be nice to know exactly what class-S technology works in the non-simply laced case. In particular, is there any relation of these theories to Hitchin systems? The answer is unclear, though there is one obvious subtlety. The Lie algebra type of the Hitchin system determines the possible scaling dimensions under the $\mathbb{C}^*$ action of the parameters of the Hitchin base. From our results, we would expect to find the integrable system describing the Coulomb branch of the $(E_6)_6$ theory which has one Coulomb branch parameter of dimension three, as a $C_2$ type Hitchin system. However, the allowed scaling dimensions for a $C_2$ type Hitchin system are two and four. Thus, it is not clear what, if anything, Hitchin systems have to do with the VOAs not corresponding to 4d $\mathcal{N}=2$ SCFTs.

\section*{Acknowledgements}
 This work was supported in part by the National Science Foundation under Grant No.~PHY--2210562. The author thanks Jacques Distler for many helpful discussions throughout this project. The author also thanks Mario Martone for comments on a rough draft of this work.
\bibliographystyle{utphys}
\bibliography{references}

\end{document}